\documentclass[preprint]{aastex}
\def\simgr{\,\hbox{\hbox{$ > $}\kern -0.8em \lower 1.0ex\hbox{$\sim$}}\,}
\def\simle{\,\hbox{\hbox{$ < $}\kern -0.8em \lower 1.0ex\hbox{$\sim$}}\,}

\shortauthors{Thorstensen \& Halpern}
\shorttitle{X-ray Selected Cataclysmic Binaries}

\def\rxs{1RXS J045707.4+452751}
\def\swa{Swift J0732.5$-$1331}
\def\swb{Swift J0746.2$-$1611}
\def\axa{AX J1740.2$-$2903}
\def\igra{IGR J18173$-$2509}
\def\igrb{IGR J18308$-$1232}
\def\axb{AX J1853.3$-$0128}
\def\igrc{IGR J19267+1325}
\def\igrd{IGR J19552+0044}
\def\swc{Swift J2218.4+1925}

\def\ro{{\it ROSAT\/}}
\def\int{{\it INTEGRAL\/} IBIS}
\def\chandra{{\it Chandra}}
\def\xmm{{\it XMM--Newton}}
\def\sw{{\it Swift\/}}

\begin{document}
\title{Optical and X-ray Studies of Ten X-ray Selected 
Cataclysmic Binaries\footnote
{Based on observations obtained at the MDM Observatory,
operated by Dartmouth College, Columbia University, 
Ohio State University, Ohio University, and the University of Michigan.}
}
\author{John R. Thorstensen}
\affil{Department of Physics and Astronomy, Dartmouth College,
Hanover NH, 03755, USA}

\author{Jules Halpern}
\affil{Columbia Astrophysics Laboratory,
Columbia University,
550 West 120th Street, 
New York, NY 10027, USA}

\begin{abstract}
We report on ground-based optical
observations of ten cataclysmic binaries that
were discovered through their X-ray emission.
Time-resolved radial velocity spectroscopy 
yields unambiguous orbital periods for eight 
objects and ambiguous results for the remaining two.
The orbital periods range from 87~min to 9.38~hr.
We also obtained time-series optical photometry for
six targets, four of which have coherent pulsations.
These periods are 1218~s for \rxs, 628~s for \axa, 477~s
for \axb, and 935~s for \igrc.
A total of seven of the sources have coherent oscillations
in X-rays or optical, indicating that they are intermediate polars
(DQ Herculis stars).  Time-resolved
spectroscopy of one object, \swc, shows that it is an 
AM Herculis star, or polar, and \igrd\ may also be in that
class.  For another object, \swb,
we find an orbital period of 9.384 hr and detect the 
spectrum of the secondary star.  The secondary's
spectral contribution implies a distance of 
900 (+190, $-$150)~pc, where the error bars are estimated using
a Monte Carlo technique to account for correlated uncertainties.
\end{abstract}

\keywords{
cataclysmic variables --- X-rays: binaries --- stars: individual
(\rxs, \swa, \swb, \axa, \igra, \igrb, \axb, \igrc, \igrd, \swc)
}

\section{Introduction}

Cataclysmic variables (CVs) are accreting binaries in which
a late-type, usually dwarf star, donates mass via Roche-lobe overflow
to a white dwarf (WD).  In X-ray surveys, the occurrence of
different classes of CVs is correlated with the magnetic
properties of the WD.  At hard X-ray energies ($>10$~keV),
the population of CVs is dominated by magnetic systems, principally
intermediate polars (IPs, or DQ Herculis stars),
but also polars (AM Herculis stars).
In polars, the magnetic field is strong enough to channel
matter directly from the companion, along a narrow stream,
onto the magnetic pole(s) of the WD.  In the IPs, the magnetic
field is not strong enough to prevent the formation of an
accretion disk, but the disk terminates at a magnetospheric
boundary; here again, matter is channeled onto the magnetic pole(s)
of the WD.  An IP is distinguished by its WD spin period,
evident as a coherent oscillation in X-ray or optical emission
from the rotating hot spot, at a shorter period
than the orbital period of the binary.  Sometimes the beat
period between the spin and orbit periods is seen, due to reprocessed
emission.  In polars,
the WD rotates synchronously with the binary orbit, or nearly so
in the case of the few asynchronous polars (APs).  Polars are
also characterized by strong optical polarization, unlike the IPs.

X-ray spectra of CVs are correlated with their magnetic properties.
The hard X-ray properties of magnetic CVs were reviewed recently by
\citet{sca10}.  Their thermal bremsstrahlung
X-ray emission presumably originates from
the gravitational energy radiated in the shocked accretion column just
above the surface of the WD, which accounts for the dominance of
this class in hard X-ray ($>15$~keV) surveys, such as those conducted by
the \sw\ Burst Alert Telescope (BAT) and \int.
The \sw /BAT 70-month hard X-ray survey \citep{bau12}
lists 55 CVs, of which
41 are magnetic: 31 IPs and 10 polars.  Of the 
33 CVs in the the fourth IBIS/ISGRI catalog \citep{bir10},
25 are IPs and four are polars.
The IPs outnumber the polars in hard
X-ray surveys, presumably because of their higher accretion rates,
and also because in polars the accretion may be more
``blobby,'' depositing some fraction of the energy directly onto the
surface of the WD.  The later hypothesis is motivated by
observed blackbody radiation from polars
with temperatures of tens of eV.  Thus, the polars sometimes
have two-component, soft and hard spectra, which explains why
many were discovered in the soft X-ray survey
of the \ro\ satellite.
 
\begin{figure}
\epsscale{0.95}
\plotone{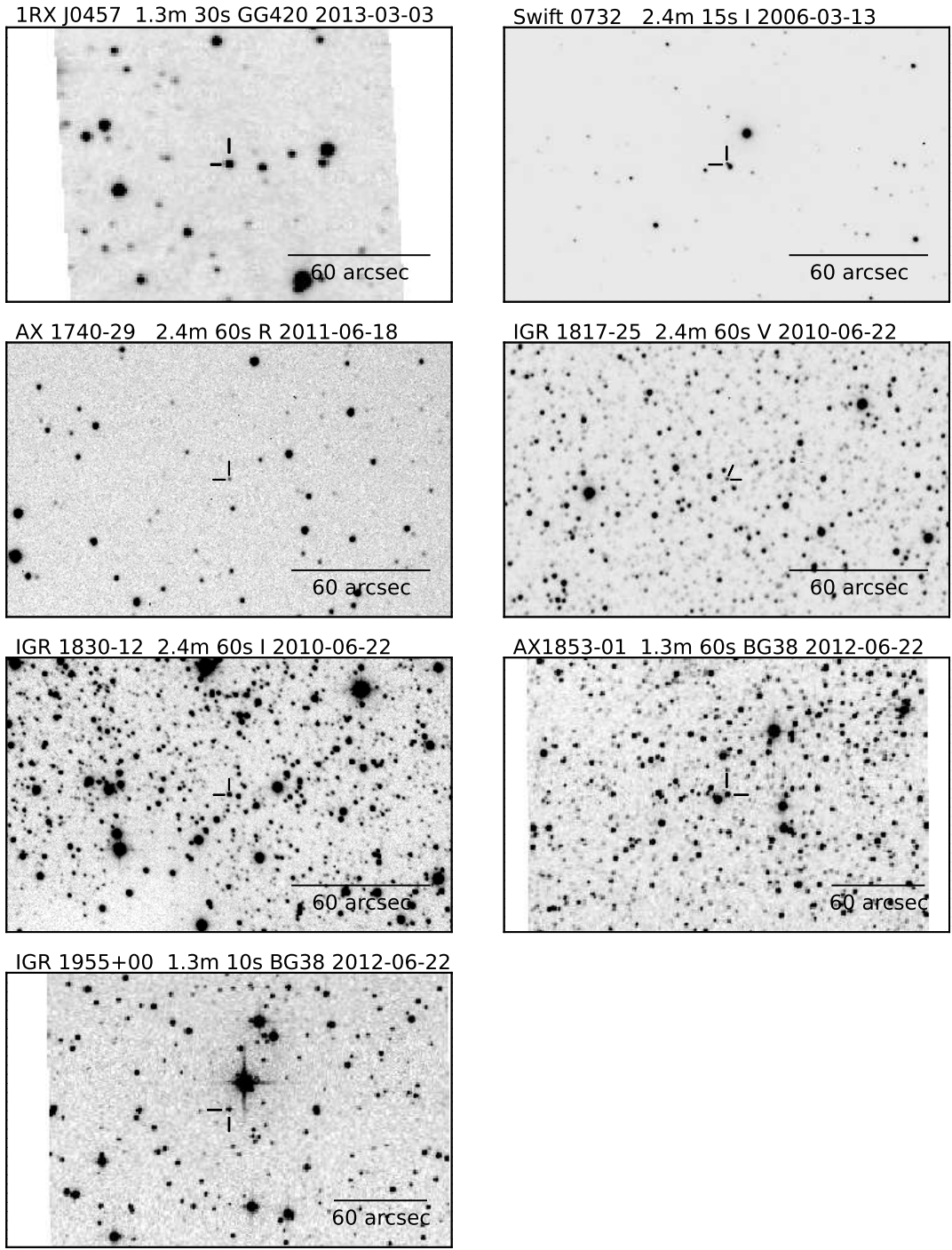}
\caption{Finding charts for some of the objects.  North is up,
east is to the left, and the scale is indicated.  The label on 
each Figure gives the source of the image used, and the tick marks
are accurately aligned with the coordinates in Table~\ref{tab:objects}.  
Shortened names are used; ``Swift 0732'' is V667 Puppis.  The chart
for \rxs\ supersedes the incorrect chart in \citet{mas10}.
}
\label{fig:charts}
\end{figure}

Here we report on ground-based optical studies of ten 
CVs discovered by either {\it ASCA\/}, \sw /BAT, or \int.
As one might expect from the discussion above, nearly all these
show evidence of being magnetic, such as relatively strong
He~II~$\lambda$4686 emission caused by X-ray photoionization,
or coherent oscillations.
Section \ref{sec:techniques} 
gives details of the observations and their analysis, and   
Section~\ref{sec:individuals} describes the results for 
the individual objects.   The objects observed
are listed in Table \ref{tab:objects}, along with accurate
celestial coordinates, approximate magnitudes, and spin periods
where observed.  Figure~\ref{fig:charts} gives
finding charts of the objects for which we have
our own direct images.  Table \ref{tab:velocities} lists
individual radial velocities, and Table \ref{tab:parameters} gives parameters
of the best-fit sinusoids.  Figures~\ref{fig:montage1} and
\ref{fig:montage2} show mean spectra, radial velocity
periodograms, and folded radial velocities for all but
two of the objects.  We also present discovery
data from the \sw\ X-ray Telescope (XRT) and Ultraviolet/Optical
Telescope (UVOT), as well as time-series
photometry and period analysis of detected pulsations
in optical and X-ray.  Our conclusions are summarized
in Section~\ref{sec:conclusions}.

\section{Equipment and Techniques}
\label{sec:techniques}

All our optical data are from the MDM Observatory,
which comprises the  1.3m McGraw Hill telescope
and the 2.4m Hiltner telescope, on Kitt Peak, Arizona.
High cadence photometry was carried out on six
targets to search for spin periods that would originate
from an IP.  Five of these
stars were observed with the the 1.3m, and a sixth with
the 2.4~m.  All ten of our radial velocity studies to
search for the orbital periods used the 2.4m.

\subsection{Time Series Photometry}

For the time-series photometry in 2008--2012, we 
used the thinned, back-illuminated SITe CCD ``Templeton,''
a $1024\times1024$ chip with $24\,\mu$m pixels.  In order to minimize
readout time, the CCD was windowed and binned $2\times2$ to
cover $4.\!^{\prime}3\times4.\!^{\prime}3$ with
$1.\!^{\prime\prime}0$ pixels.  Exposure times ranged from 10--30~s
with a dead-time between exposures of only 3.6~s.  For relatively
blue stars we used a Schott BG38 filter to maximize throughput,
while a reddened star required an $R$ filter for optimal signal.
Standard reductions used bias frames and sky flats, and
aperture photometry was performed with the IRAF\footnote
{IRAF is distributed by the National Optical
Astronomy Observatory, which is operated by the Association
of Universities for Research in Astronomy (AURA) under
cooperative agreement with the National Science Foundation.
} 
routine {\tt phot} and a single comparison star that was
2--3 magnitudes brighter than the variable.  \citet{lan92}
standard stars were used to calibrate the $R$-band frames.
For the BG38 filter, an approximate magnitude was derived
by averaging the $B$ and $R$ magnitudes of the comparison
star in the USNO-B1.0 catalog \citep{mon03}.

Starting in 2013 we obtained time-series photometry with the 1.3-m
telescope and an Ikon DU-937N CCD camera built by Andor
Technology PLC.  This 
integrated unit employs a thinned, backside-illuminated
frame-transfer CCD with 13~$\mu$m square pixels and
a $512\times512$ active area that subtends $140''$ at the 
1.3-m.  With $4\times4$ binning,
the pixel scale was $1.\!^{\prime\prime}1$~pixel$^{-1}$.  
Individual exposures were 10--30 seconds.  We used Schott glass
GG420 or BG38 filters, or a $V$-band filter.
The dead time between exposures was only 11.92 milliseconds.
The sensor was thermoelectrically cooled to $-50$ C for all 
observations.  The Andor {\it Solis} software recorded
each observation sequence as a data cube, giving only 
a start time and an inter-frame interval rather than 
a separate start time for each image; extensive tests
showed that the nominal inter-frame interval was 
sufficiently accurate for our purposes.
To reduce the data cubes, we
subtracted median bias images and median dark images (scaled
by exposure times), and divided by normalized median images
of the twilight sky in the appropriate filter.  
We again measured magnitudes using  
{\tt phot}, and referred these to a comparison star
in the field.  For the GG420 filter, this was the
$R$ magnitude of the comparison
star in the USNO-B1.0.  

The comparison star light curves often
showed variations due to clouds, but these were for the most
part thin enough that the differential photometry was
usable.  Sequences on individual stars ranged from 2.4--6.5 hours.
Period searches used a standard power spectrum analysis on
these evenly sampled light curves.  In two cases a slow
trend was fitted out before calculating the power spectrum
of the ``detrended'' light curve.

\subsection{Spectroscopy}

Nearly all of our radial velocity studies on the 2.4m used the modular
spectrograph.  The detector was one of two thinned SITe
CCDs, either ``Echelle'', a $2048\times2048$ chip that
covered $4300-7500$~\AA\ at 2.0~\AA\ pixel$^{-1}$, or
``Templeton'', the aforementioned $1024\times1024$ chip that covered
$4660-6740$~\AA\ with the same dispersion.  Wavelength
calibrations were from Hg, Ne, and Xe comparison lamps
taken during twilight, and we used the night-sky
$\lambda5577$ feature to track spectrograph flexure
during the night.  When it was clear, we observed
flux standards during evening and morning twilight 
to calibrate the instrument response, but 
photometric conditions did not always prevail for our
program-star observations, and 
seeing variations caused an
an unknown fraction of the light to be lost on the jaws of the 
$1^{\prime\prime}$ (projected) slit.  Experience
suggests that the fluxes in our average spectra are 
typically accurate to 
$\sim 20\%$.  In addition, our modular
spectrograph setup sometimes produces unrealistic
continua, which fortunately tend to average out over many 
exposures.  Our slit was oriented north-south by
default -- parallel to atmospheric dispersion
on the meridian.   At large hour angles and zenith
distances,
we rotated the instrument
to orient the slit close to 
the parallactic angle\footnote{The most recent version
of the {\it JSkyCalc} program, available at JRT's
website, includes a 
parallactic angle optimization tool.  Note that the 
dispersion grows as $\tan z$, rather than $\sec z$, so there
is no large ``flat spot'' around the zenith as there is
with extinction.}.  This was
often needed, because we were trying 
to find orbital periods; observations far from the
meridian are required in order to suppress ambiguities
in the daily cycle count.  (For one target, \swa, 
we could not optimize the slit position angle because we needed
to avoid a nearby brighter star.)

Our 2012 December velocities of \igrd\ are from the
modular spectrograph mounted on the 1.3m telescope.
The setup and procedures were identical to those 
used at the 2.4m, except that we
did not rotate the instrument to the parallactic angle.
The 1.3m and 2.4m telescopes are both $f7.5$, so the 
image scale of the 1.3m is proportionally smaller than that
of the 2.4m.  Because the same physical slit size was used 
on both telescopes, the projected slit width was larger 
at the 1.3m.  Consequently,   
alignment with the parallactic angle is less critical than
with the larger telescope. 

We reduced our spectra using standard IRAF
and Pyraf routines for
bias subtraction, flatfielding, wavelength
calibration, and flux calibration.  To extract
one-dimensional spectra, we used an implementation
of the algorithm published by \citet{horne}.   We
measured emission-line radial velocities using
convolution algorithms described by \citet{sy80} and
\citet{shafter83}.  When a cool star was detectable
in the spectrum, we used the \citet{kurtzmink} 
implementation of the \citet{tondav} cross-correlation
to measure its velocity; for the template, we used 
a velocity-compensated sum of 76
observations of late G and early K-type IAU
velocity standards.  To measure synthetic 
magnitudes from our spectra we used the
\citet{bessell} tabulation of the $V$ passband
and the IRAF {\it sbands} task.

The main purpose of our spectroscopy was to determine the
periodicity in the radial velocities (which is 
identical to the orbital period $P_{\rm orb}$ 
in nearly all cases).  We searched for periods 
in the time series using a ``residual-gram'' algorithm
\citep{tpst}.  The observability constraints imposed
on observations from a single site inevitably lead
to some ambiguity in the daily cycle count;
when these appeared significant, we used the 
Monte Carlo method of \citet{tf85} to assess the 
robustness of our choice of orbital period.
We use the ``discriminatory power" statistic from that test, which
is the fraction of Monte Carlo trials that result in the correct period
being chosen, and not an alias, by virtue of having the smallest residuals.

\section{Individual Objects}
\label{sec:individuals}

Targets for this study were selected via a variety of methods
that were likely to yield magnetic CVs.
These include followup \xmm\ observations showing periodicities
from {\it ASCA\/} sources in the Galactic plane, \sw\ BAT sources
for which pointed observations with the XRT found rapidly varying
X-rays and a relatively bright or blue optical counterpart,
and \int\ sources already identified spectroscopically as CVs.
In this section, we detail our observations and results on the
individual objects.

\subsection{\rxs} 

\begin{figure}
\epsscale{0.95}
\vspace{-1.0in}
\centerline{
\includegraphics[scale=0.85,angle=0]{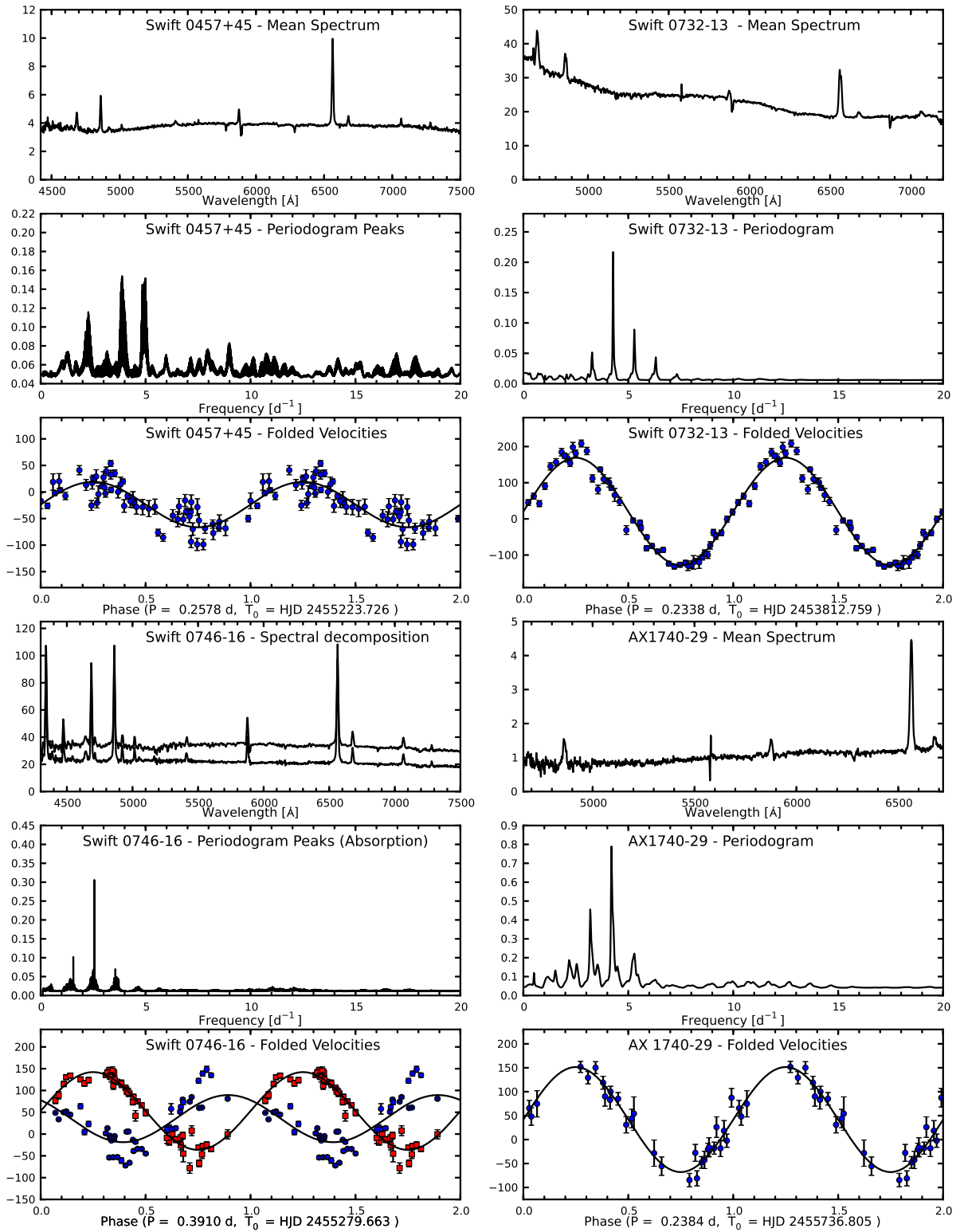}
}
\vspace{0.5in}
\caption{Mean spectra, periodograms, and folded velocity curves for 
four of the objects.  The units of the vertical axes of the spectra are
$10^{-16}$ erg cm$^{-2}$ s$^{-1}$ \AA$^{-1}$; for the 
periodograms, the axis is unitless ($1 / \chi^2$); and the 
radial velocities are in km s$^{-1}$.  
In the velocity curves, all data are repeated on an extra cycle for 
continuity, the uncertainties shown are estimated from 
counting statistics, and the solid curves show the 
best-fitting sinusoids.  In the spectrum of \swb,
the lower trace shows the result of subtracting a K4V-type spectrum,
scaled to $V = 16.36$.   Circles (blue in on-line version) 
are emission-line velocities
and squares (in the case of \swb; red in on-line version) are 
absorption-line velocities of the secondary star.
}
\label{fig:montage1}
\end{figure}

The IR counterpart of this \ro\ X-ray source was identified
by \citet{kap06} from a \chandra\ image, but not classified.
It was detected in the \sw\ BAT survey \citep{vos10,cus10a},
and by \int\ \citep{kri10}, then classified as a CV when
\citet{mas10} obtained a spectrum of the optical counterpart.
\citet{mas10} give coordinates and a finding chart for the wrong star;
Fig.~1 and Table 1 give the correct identification.
Our spectrum 
(Fig.~\ref{fig:montage1}) appears 
nearly identical to theirs, but with a slightly lower
continuum level and better signal-to-noise ratio.
In agreement with \citet{mas10} we see He~II $\lambda4686$ 
at less than half the strength of H$\beta$; our 
signal-to-noise reveals He~II $\lambda$5411 as well, 
with an emission equivalent width (EW) of $\approx 1.7$ \AA .
We detect the diffuse interstellar bands (DIBs) at 
$\lambda5780$ (0.7 \AA\ EW) and $\lambda6283$
(1.0 \AA\ EW).  \citet{YuanLiuDIB}, in a review
of SDSS spectra, find on average that
$EW(5780)=0.61\times E_{B−V}$ and $EW(6283) =
1.26 \times E_{B−V}$, with considerable scatter, so these
band strengths are roughly consistent with $E(B-V)\approx1.0$.
The \citet{schlegel98} extinction map gives $E(B-V)=1.05$
at this location (or 0.90 if one adopts the 14\% correction
recommended by \citealt{sch11}), in excellent agreement with the 
DIB estimate.
Our spectrum implies $V=17.5$; taking $A_V/E(B-V)=3.1$ 
gives an extinction-corrected $V=14.7$.  The object
lies at Galactic coordinates $l=160.\!^{\circ}97,\ b=+1.\!^{\circ}52$,
so the sizeable reddening is not unexpected.


The object appears to be a novalike variable of 
some kind.  At a distance of 1~kpc, it would
have $M_V\sim+4$, comparable to other
novalike variables \citep{warner95}. 
\citet{mas10} estimate $A_V\sim0$ and
$d\sim500$ pc for this object (their Table 2); the 
substantial reddening we find, and the similarity
to a novalike variable, suggest that they underestimate
both these quantities.

Our H$\alpha$ emission velocities (Fig.~\ref{fig:montage1}) did not yield
an unambiguous orbital period, but we do detect a 
significant periodicity consistent with either 
$P_{\rm spec}\approx0.258$ d (3.87~cycle~d$^{-1}$) 
or $\approx0.200$~d (4.98~cycle~d$^{-1}$).  These
are daily cycle count aliases.  We also do not
know how many cycles elapsed between our observing
runs.  We are at least confident that
the orbital period is longward of the 2--3~hr ``gap''
in the CV period distribution.  

Figure~\ref{fig:0457pulse_jan} shows a 6.8-hour photometric sequence 
we obtained with the Andor camera on 2013 January~4, 
using a GG420 filter.  The magnitude scale is calibrated
to the $R$ magnitude of the comparison
star from the USNO-B1.0 catalog.
The main period detected is 
$1208\pm12$ s, which is likely to be the rotation 
period of a magnetized white dwarf (or possibly half the
period).  Detrending the data with a 0.258-d
period improves the power spectrum slightly.  The night
was mostly clear, though the comparison
star showed evidence of occasional light clouds.  The 1208-s
period was not detected on two subsequent nights,
but our observations were shorter then, and we used
a BG38 filter, which proved to be less sensitive than
the GG420.

We obtained additional time series with the same setup on
2013 March 1--3 (Fig.~\ref{fig:0457pulse_mar}). These displayed
the periodic signal very clearly, and a coherent fit
to the three consecutive nights yields $P=1218.7\pm0.5$~s,
which we adopt as the most precise value.
There is no evidence from the folded light curves
that the true period is twice this value.
\citet{mas10} suggested that the system is 
not magnetic, on the basis of its relatively weak He~II 
emission, but the detection of a coherent modulation
shows that it actually is magnetic.  We have examined the
existing \chandra\ and \sw\ XRT data on \rxs,
but they are too sparse to reveal the period.

\subsection{\swa = V667 Puppis}
\label{sec:swift0732}

A flurry of discovery surrounded this source in 
early 2006.
First, \citet{ajello06} detected it using the 
\sw\ Burst Alert Telescope (BAT) and suggested
it was identical to a \ro\ source coincident with
a $B \sim 15$ mag star.  \citet{masetti06} found
emission lines, confirming the identification; 
\citet{patterson06} found a periodic modulation at
512.42(3) s, and pulsations at the same frequency were 
soon found by \citet{wheatley06} using the \sw\ XRT, confirming
the IP nature of the source.  Further
confirmation of the spin period was found by 
\citet{but07} in {\it RXTE\/} data.
\citet{patterson06} also obtained a spectrum showing a 
strong G-type stellar contribution to the spectrum, but were
unable to detect the ellipsoidal modulation that would be
expected if the G-star were the secondary.  
The puzzle of the missing ellipsoidal modulation was resolved when 
\citet{marsh06} found the purported
counterpart to be a close optical pair, with the fainter
of the two stars being the bluer, pulsating object. 
\citet{torres06} found significant emission line
velocity variations, but were unable to determine a 
period.  We then obtained more extensive velocities and found the
0.2335-d period reported briefly in \citet{thorstensen06}.
We present these observations in greater detail here.

The spectrum (Fig.~\ref{fig:montage1}) shows relatively
weak emission, with He~II~$\lambda 4686$ around the same
strength as H$\beta$.  There is a broad bump in the 
continuum between 5000~\AA, and 6200~\AA, but we cannot
be sure this is real; the strong upsweep into the blue
also appears unphysical.  Because of the crowding
star, we were uanble to follow our usual practice of
keeping the slit near the parallactic angle; instead,
we used slit position angles between $-36^{\circ}$ and $-44^{\circ}$.
Close examination of the spectrum reveals weak,
stationary absorption features around
$\lambda5175$, which we believe are from the
crowding star; the crowding star may also be 
responsible for the continuum bump.  

The H$\alpha$ emission line velocities show a 
very clear modulation at 5.61~hr.  
Figure~\ref{fig:swift0732trail} shows a greyscale 
phase-averaged spectrum synthesized from our spectra.
To prepare this, we (1) rectified our
spectra to the continuum, (2) hand-edited the individual spectra
to remove radiation events and other artifacts,
(3) constructed each line of the image from a weighted
average of spectra close to the phase represented by that
line\footnote{The weighting function used was a Gaussian in phase, with 
a full-width at half-maximum of $\sim 0.05$ cycles, truncated $\pm 0.05$ cycles from the 
central phase.}, and (4) stacked the lines to form the two-dimensional image.  
There is a rather faint double-peaked profile, with
a modest velocity amplitude, and a stronger 
component that undergoes an ``S-wave'' modulation
between the velocities of the peaks.  This may 
arise in a hot spot, where the mass-transfer stream
strikes the accretion disk.  On the other hand, the sizeable radial
velocity amplitude is reminiscent of an AM~Her
star (or polar), in which there is no accretion disk
and therefore no hot spot.
In polars, the accretion column
produces large infall velocities, and hence
strong velocity modulation, but the line profile does
not support this interpretation. 
AM~Her stars tend to show broad, asymmetric line wings, 
which move with the orbit (as in Swift J2218,
Section~\ref{sec:swift2218}).
These are not seen here, so 
this appears not to be a polar.

We obtained three sets of $UBVI$ images on 
2006 March 13 UT, with the 2.4m telescope.
Conditions were photometric, and the 
seeing was $\approx1^{\prime\prime}$, which was
good enough to resolve the X-ray source from the
nearby star, as shown in an $I$-band image in Figure~\ref{fig:charts}.
To measure magnitudes, we used the 
IRAF implementation of DAOphot, which fits
point spread functions.  The position in 
Table \ref{tab:objects} is from this fit.
For V667 Pup, we found 
$V=15.8,\ B-V=+0.3,\ U-B=-0.7$, and
$V-I=+0.6$, with fluctuations between
measurements of $\sim0.05$~mag.  The 
companion averaged $V=14.29$, $B-V=+0.66$, 
$U-B=+0.13$, and $V-I=+0.75$, with somewhat 
smaller fluctuations.  

\subsection{\swb}  

This is an unclassified source in the \sw\ BAT survey
\citep{vos10,cus10a}, where it was associated with the \ro\ source
1RXS J074616.8$-$161127.  It was also detected by the {\it BeppoSAX\/}
Wide Field Camera \citep{cap11}.
We identified it with a highly variable
X-ray source in several \sw\ XRT observations (Fig.~\ref{fig:swiftxrt}),
as a well as with a UV-bright counterpart in the \sw\ UVOT
(Fig.~\ref{fig:swiftuvot}).  The \sw\ observations are too sparse
to reveal any periods.  We are not aware of any prior optical
work on \swb.

Its spectrum (Fig.~\ref{fig:montage1}),
includes a significant contribution from a 
late-type secondary star as well as the typical
emission lines.  The secondary's
velocities give period near 9~hr, which we refined 
over several observing runs to 0.39101(1)~d.  The
emission-line velocities are much less steady, but
corroborate the absorption-line period (and eliminate
the possibility that the absorption lines are from 
a distant companion to the binary, or chance superposition).

By scaling and subtracting
spectra of stars classified by \citet{keenan89}, and 
searching for good cancellation of the secondary's
features, we determined the secondary's spectral type to 
be K4 $\pm$ 2 subclasses, and quantified its contribution
to the spectrum.  We then computed a distance based on the 
following procedure, which is described in more detail
by \citet{thorstensen12} and \citet{peters05}.
(1) The Roche-lobe constraints and $P_{\rm orb}$ determine the 
physical size of the secondary (with a weak dependence
on the secondary's assumed mass). (2) The secondary's spectral type
determines its surface brightness.  (3) 
The spectral decomposition gives an apparent magnitude
for the secondary's contribution.  (4) We correct for
reddening, and determine the distance.  

All of these steps involve uncertainties; a crude
estimate of the uncertainty in the final distance can
be had by propagating the errors in the standard manner,
and that is the procedure we have followed in the past.
This is not entirely satisfactory, 
because some of the parameters can be 
correlated.  Most importantly, metal lines are
stronger at later spectral types, so normalization
of the secondary star's spectral contribution -- which is
estimated by looking for the best subtraction of the line
features -- tends to be correlated with the 
spectral type.  We therefore implemented a Monte Carlo
procedure, as follows.  First, for each assumed 
secondary spectral type, we defined a permitted range
of secondary-star $V$ magnitudes, based on our decompositions.
We also assumed a normally-distributed 0.2 mag uncertainty
in our spectrophotometric normalization.   In this case,
acceptable
decompositions could be found for K2 through K6; we 
assigned  discrete probabilities of 0.3 to K4, 0.2 to K3 and K5, and
0.15 to K2 and K6.  For the surface brightness
at each spectral type, we used an analytical fit
to surface brightness vs. spectral type data
presented by \citet{beuermann99}, 
and assumed a $\pm0.2$ mag range of uncertainty in this
calibration.
For the secondary star mass we took $M_2=0.65\pm0.2$,
based on the \citet{bk00} evolutionary calculation; 
we took masses to be uniformly distributed within this
range and used the random masses to 
to compute the secondary's radius.  The surface 
brightness and radius then gave the absolute magnitude,
which combined with the secondary's $V$ gave an 
apparent distance modulus.  
For reddening, we used the \citet{schlegel98} maps to establish
an upper limit of $E(B-V)=0.18$, and -- since the
star is within the dust layer and evidently not
exceptionally distant -- we took the reddening to 
be $E(B-V)=0.10\pm0.05$, with an upper cutoff 
at 0.18 and a lower cutoff at 0.02\footnote{The reddening
was small
enough and uncertain enough that we ignored the 14\% 
correction recommended by \citet{sch11}.}.  With all these
ingredients in place, we computed 10,000 distances,
drawing each ingredient from its assumed distribution
every time.   The resulting distances are shown in 
Figure~\ref{fig:montecarlo}; the median and the ends of the
68\% confidence error bars give a distance of
900 (+190,$-$150) pc.  

We searched for optical pulsations in this source using the 
Andor camera on three consecutive nights in 2013 January.
The source varied  erratically by several tenths of a magnitude,
but we were unable to detect any periodicity.
The longest of the nightly light curves
is shown in Figure~\ref{fig:swiftj0746andor}.
\swb\ is a novalike variable, but whether it is also a DQ Her
star remains an open question.

\subsection{\axa}

\citet{sakano00} detected this hard source in 
an {\it ASCA\/} observation (in which they also detected another
pulsating source, AX J1740.1$-$2847).  \axa\ was
also detected by  \int\ \citep{bir10}.  
\citet{farrell10}
detected a 626-s pulse in \xmm\ data, and around the
same time \citet{malizia10}
improved the position, also with \xmm\ data, but they
gave a position that was incorrect by $25^{\prime\prime}$.
We had independently analyzed the
same \xmm\ data, finding the optical counterpart \citep{halpgott10a} 
and the pulsations, both in X-ray and then in optical \citep{halpgott10b}.
\citet{masetti12} present a optical spectrum, which 
appears similar to ours.  

We obtained time-series photometry on one night each in 2010 June and 2012 June
(Fig.~\ref{fig:photometry1}).
Because of moderate extinction to the source, only the $R$ band gave sufficient signal.
A period was clearly detected in power spectra of the light curves,
with values of $622.7\pm4.5$~s and $628.6\pm2.3$~s, respectively,
in 2010 and 2012.  These agree with
the X-ray period of $626\pm2$~s measured by \citet{farrell10},
or $623\pm2$~s in our own analysis of the same X-ray data.  The pulsed
amplitude in the optical was $\approx0.16$~mag peak-to-trough
in 2010, and $\approx0.11$~mag
in 2012.  The shape of the single-peaked light curve is similar to that
of the X-ray, except that the latter is much more highly modulated
(Fig.~\ref{fig:xraypulses}).

The spectrum (Fig.~\ref{fig:montage1}) shows
prominent H$\alpha$, with 45~\AA\ emission EW, and 
flux $\sim5\times10^{-15}$ erg~cm$^{-2}$~s$^{-1}$. 
The line is single-peaked and somewhat narrow for 
a CV, with FWHM $\approx680$~km~s$^{-1}$.  
The continuum slopes upward toward the red, suggesting 
substantial reddening, and the synthesized 
$V=19.1$ is on the faint side.  
The X-ray spectral fits of \citet{farrell10} require a
column density $N_{\rm H}\approx4\times10^{21}$~cm$^{-2}$,
which would be equivalent to $A_V\approx2$
if it is all intervening.  An absorption
feature with EW $\approx2.2$~\AA\ is present at 6283~\AA,
and there is a hint of a feature (EW $\approx0.6$~\AA)
near 5780~\AA; these are probably the DIBs.
In our reduction we correct approximately for 
telluric features, but because of the inevitably large 
airmass at this southerly declination it
is likely that the $\lambda6283$ DIB is 
blended with a telluric feature that strongly
overlaps it (\citealt{jenniskens}
show both the DIB and the telluric feature in detail).


Our emission-line velocities give a best period near 
0.237~d, or 5.7~hr (Table \ref{tab:parameters}), but
0.314~d (7.52~hr), which is 1~cycle~d$^{-1}$ lower
in frequency, cannot be entirely ruled out.  A 
Monte Carlo test gives a discriminatory
power \citep{tf85} of around 0.98 for our time series, so 
the period determination is reasonably secure.
The results presented here, especially the short orbital
period, rule out the hypothesis of Farrell et al. (2010)
that \axa\ is symbiotic binary.

\subsection{\igra}
\label{sec:igr1817}

\begin{figure}
\epsscale{0.95}
\vspace{-1.0in}
\centerline{
\includegraphics[scale=0.85,angle=0]{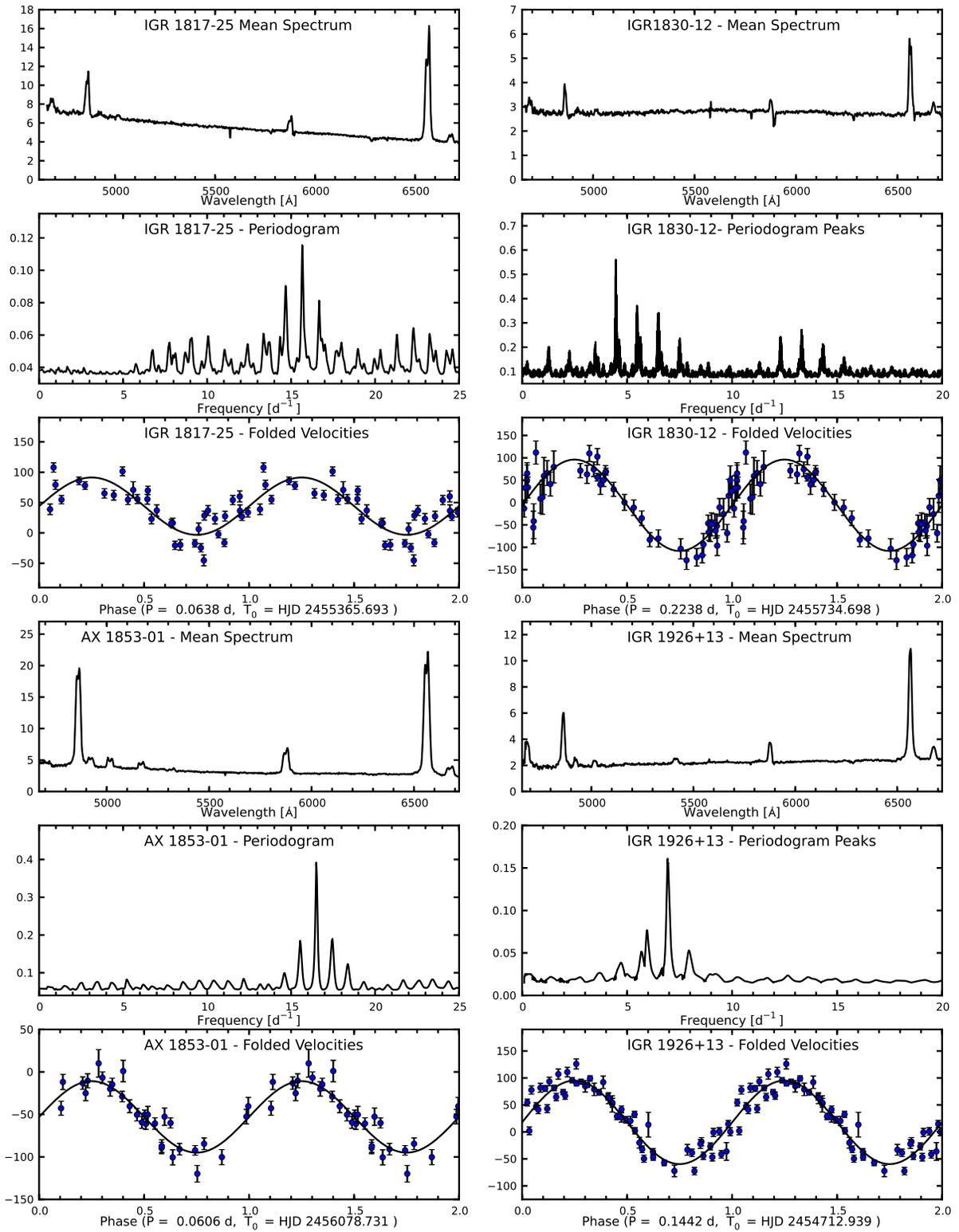}
}
\vspace{0.5in}
\caption{
Same as Figure~\ref{fig:montage1}, for four other objects.
}
\label{fig:montage2}
\end{figure}

\begin{figure}
\centerline{
\includegraphics[scale=0.82,angle=0]{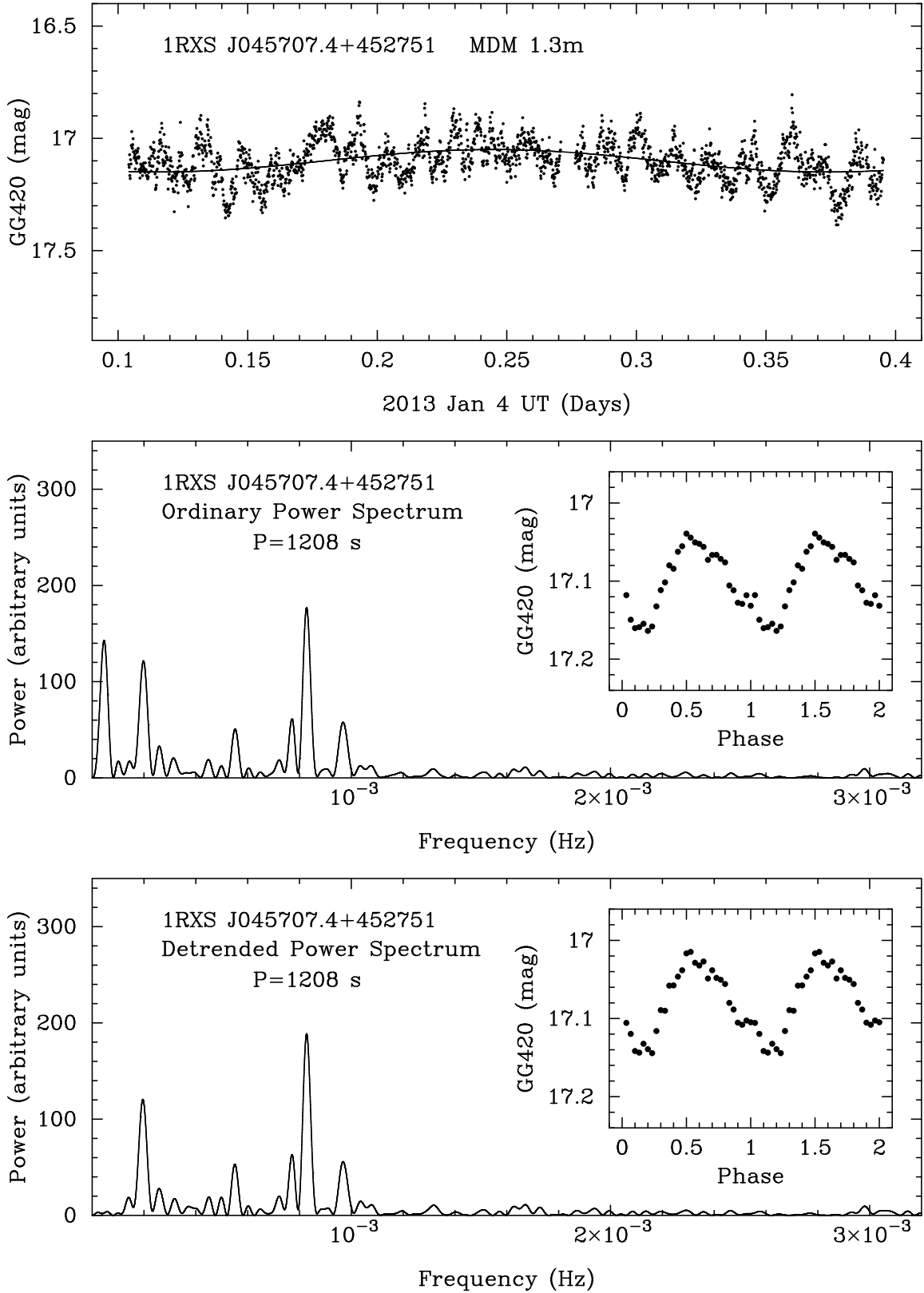}
}
\caption{Time-series photometry of \rxs, with power spectra
before and after ``detrending'' the data with a 0.258-d period.
Individual exposures are 12~s.  The rough calibration uses
the $R$ magnitude of the comparison star from the USNO-B1.0
catalog.
The insets show the mean pulse profile folded at 1208~s.
}
\label{fig:0457pulse_jan}
\end{figure}

\begin{figure}
\centerline{
\includegraphics[scale=1.,angle=0]{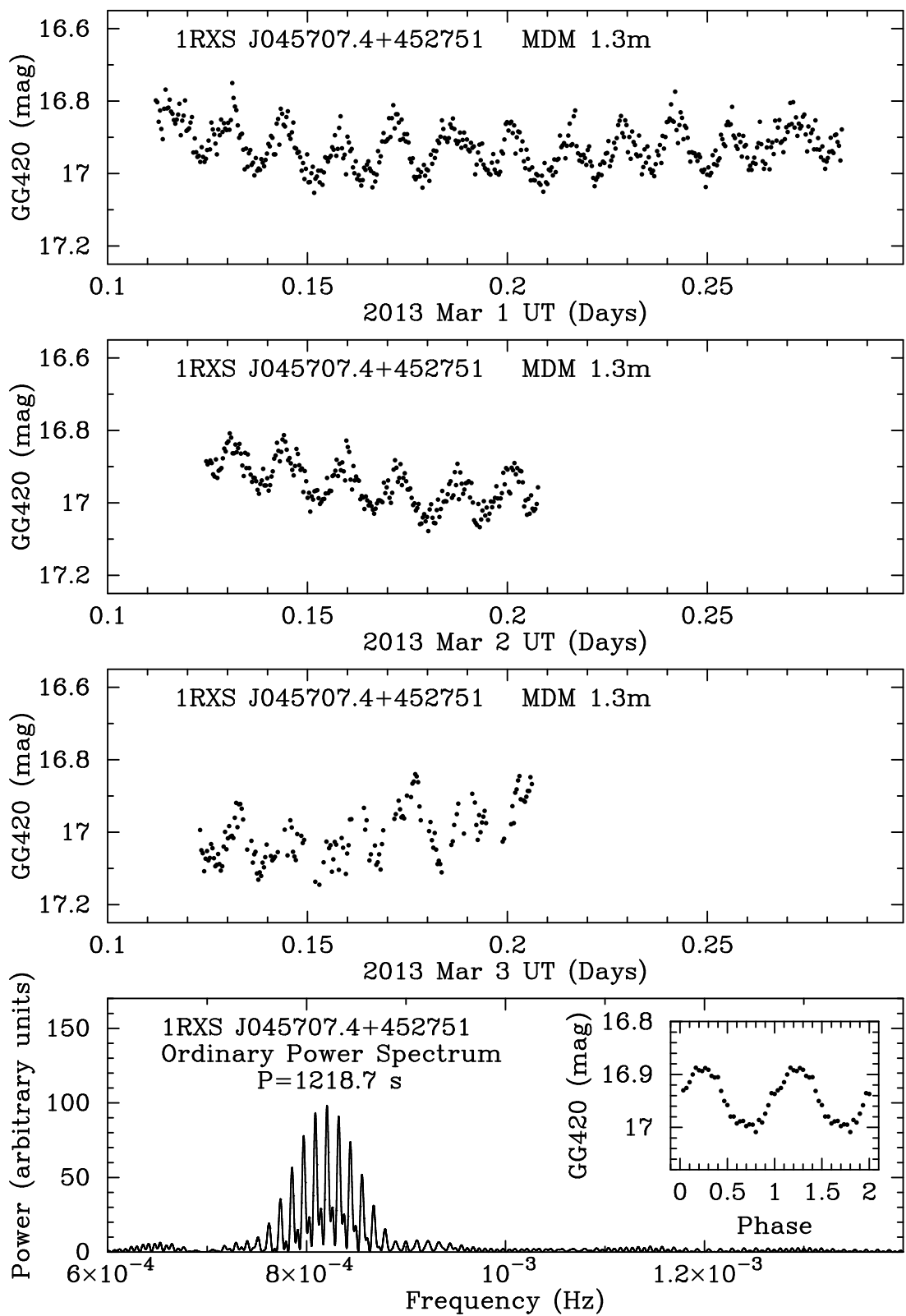}
}
\caption{More time-series photometry of \rxs.
Individual exposures are 30~s.
A coherent power spectrum of the three
consecutive nights gives a best fitted
period of $1218.7\pm0.5$~s.}
\label{fig:0457pulse_mar}
\end{figure}

\begin{figure}
\plotone{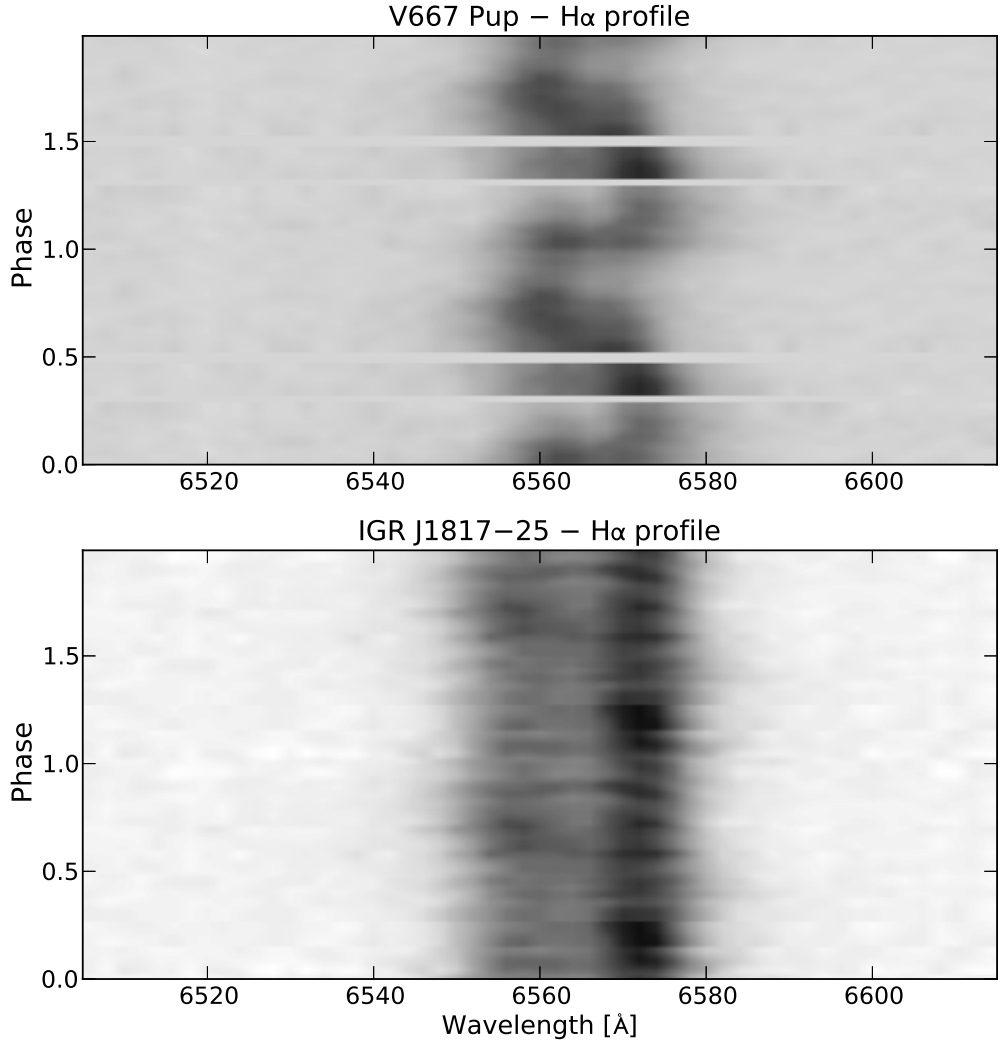}
\caption{Upper: Greyscale representation of the H$\alpha$ profile
of V667 Puppis (\swa) as a function of orbital
phase (using the ephemeris of Table~\ref{tab:parameters}).  The
data are repeated for a second cycle, and the blank 
horizontal bars are gaps in coverage.
Lower: A similar display of the H$\alpha$ profile for
\igra.
Each line in the image represents an average of spectra near that phase,
as described in Section \ref{sec:swift0732}.
}
\label{fig:swift0732trail}
\end{figure}

\begin{figure}
\vspace{-1.0in}
\hfill
\includegraphics[scale=0.85,angle=0]{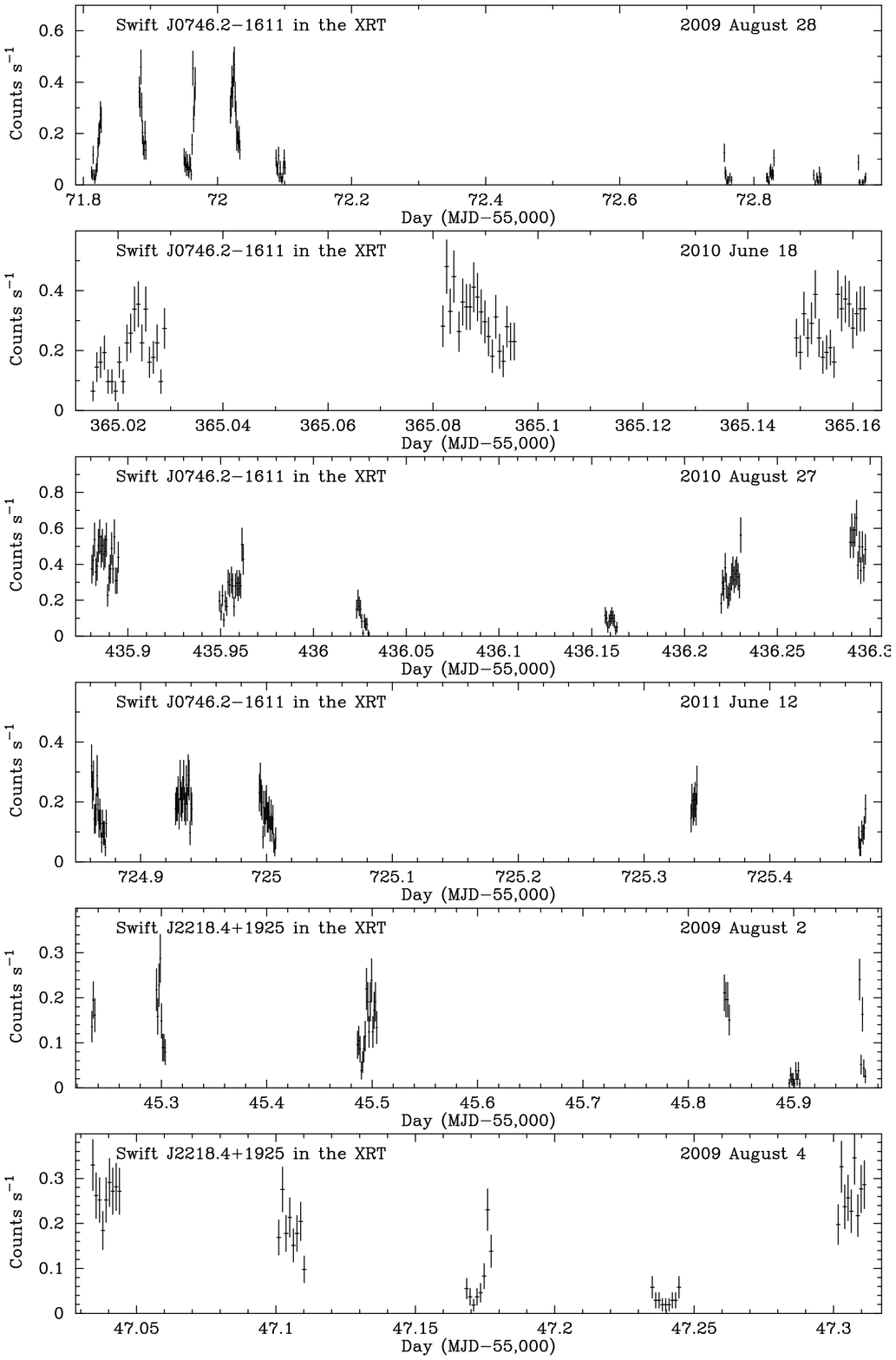}
\hfill
\vspace{-0.5in}
\caption{Light curves of \swb\ and \swc\ in the 0.3--10~keV
band from the \sw\ XRT.}
\label{fig:swiftxrt}
\end{figure}

\begin{figure}
\hfill
\includegraphics[scale=0.85,angle=0]{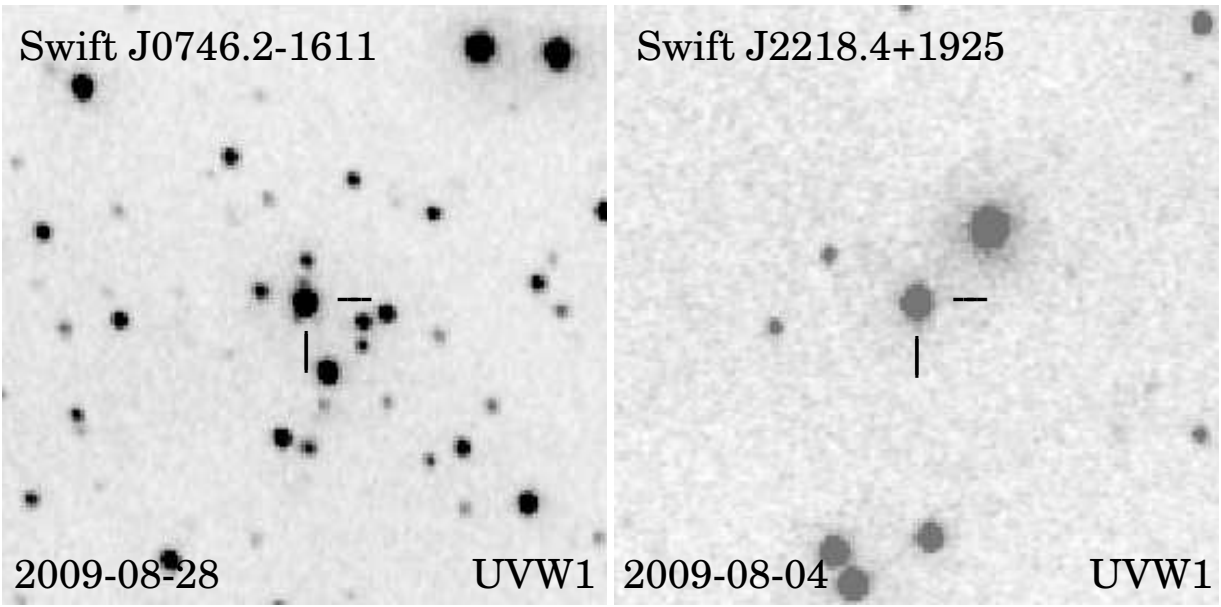}
\hfill
\caption{Finding charts for \swb\ and \swc\ in the \sw\ UVOT.
Coordinates are given in Table~\ref{tab:objects}.
Each image cutout is $3^{\prime}\times 3^{\prime}$.
North is up and east is to the left.
}
\label{fig:swiftuvot}
\end{figure}

\begin{figure}
\plotone{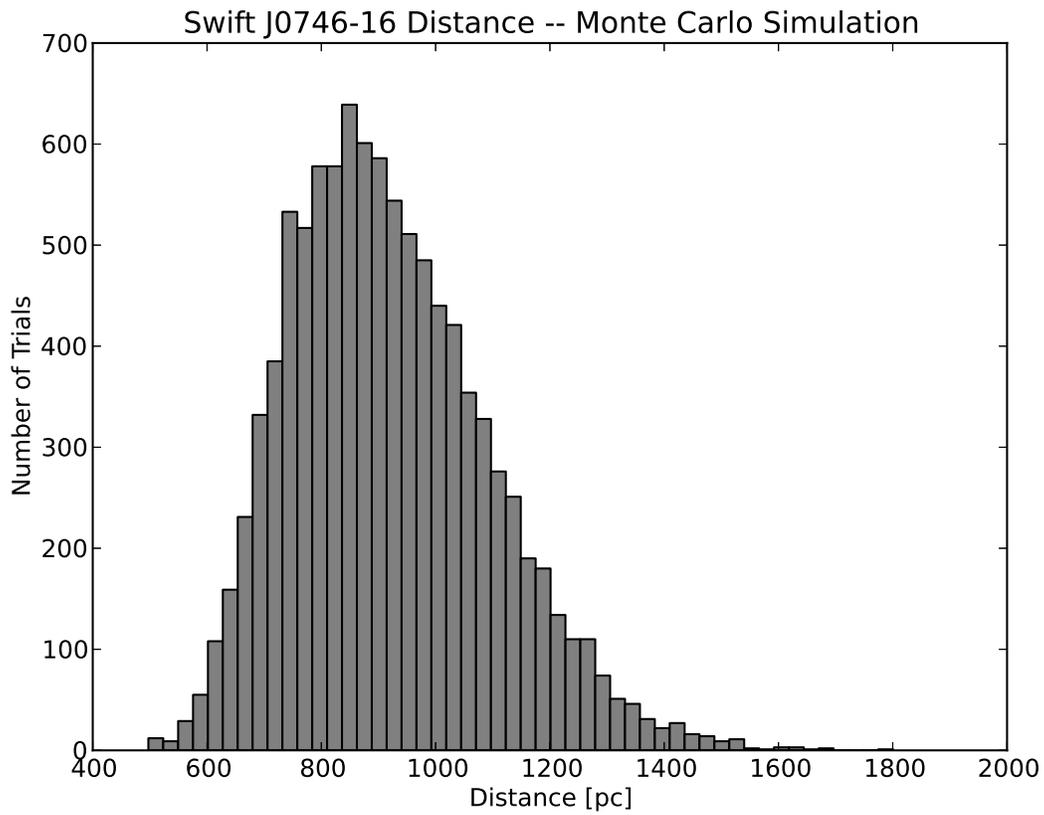}
\caption{
Distribution of distance estimates for \swb, 
from 10,000 calculations using randomized input parameters. 
}
\label{fig:montecarlo}
\end{figure}

\begin{figure}
\centerline{
\includegraphics[scale=1.,angle=0]{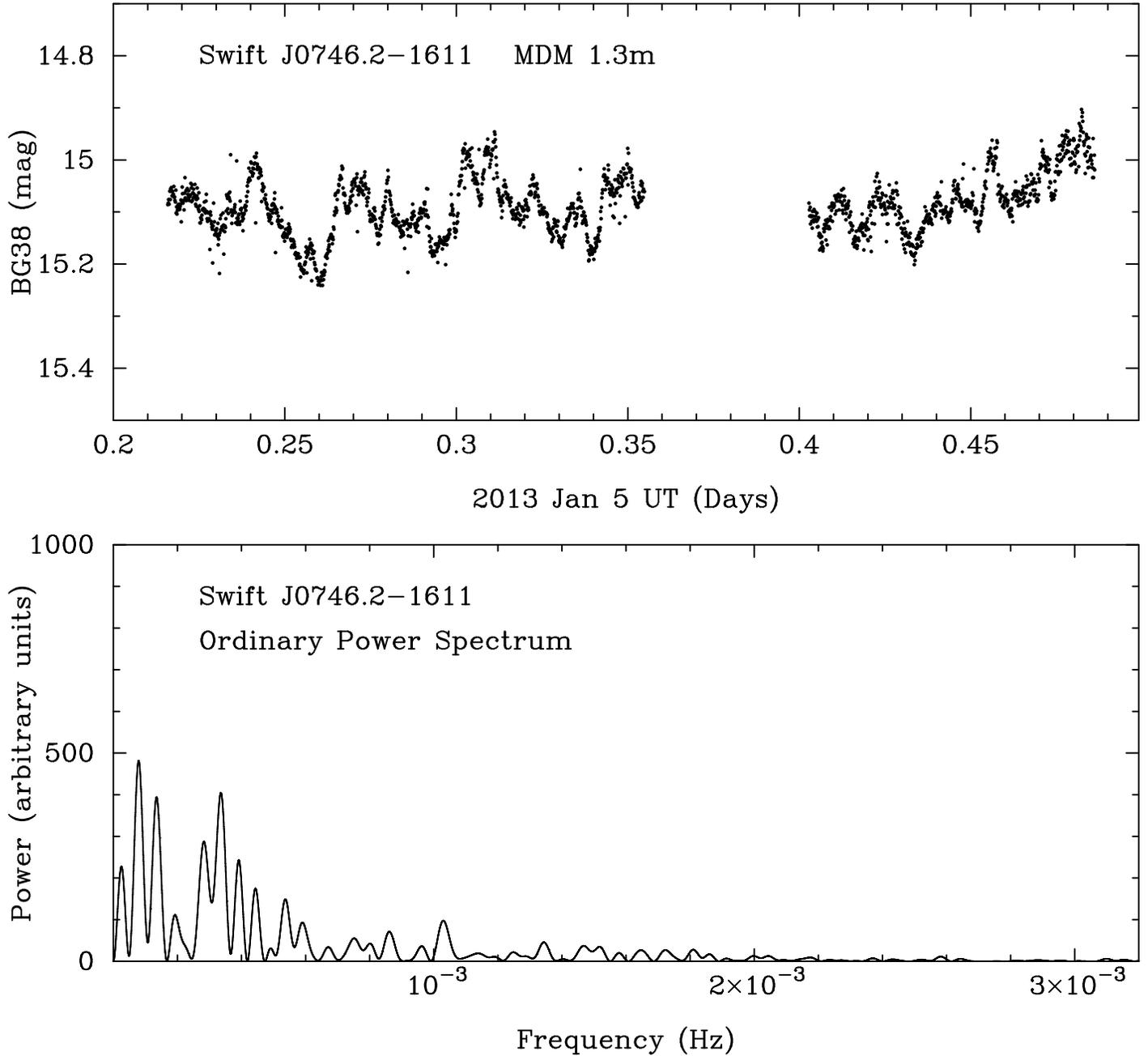}
}
\caption{Time-series photometry of \swb.
Individual exposures are 10~s.
Power spectrum analysis of this and other light curves
of this star do not reveal a period.
}
\label{fig:swiftj0746andor}
\end{figure}

\begin{figure}
\vspace{-2.0in}
\centerline{
\includegraphics[scale=1.0,angle=0]{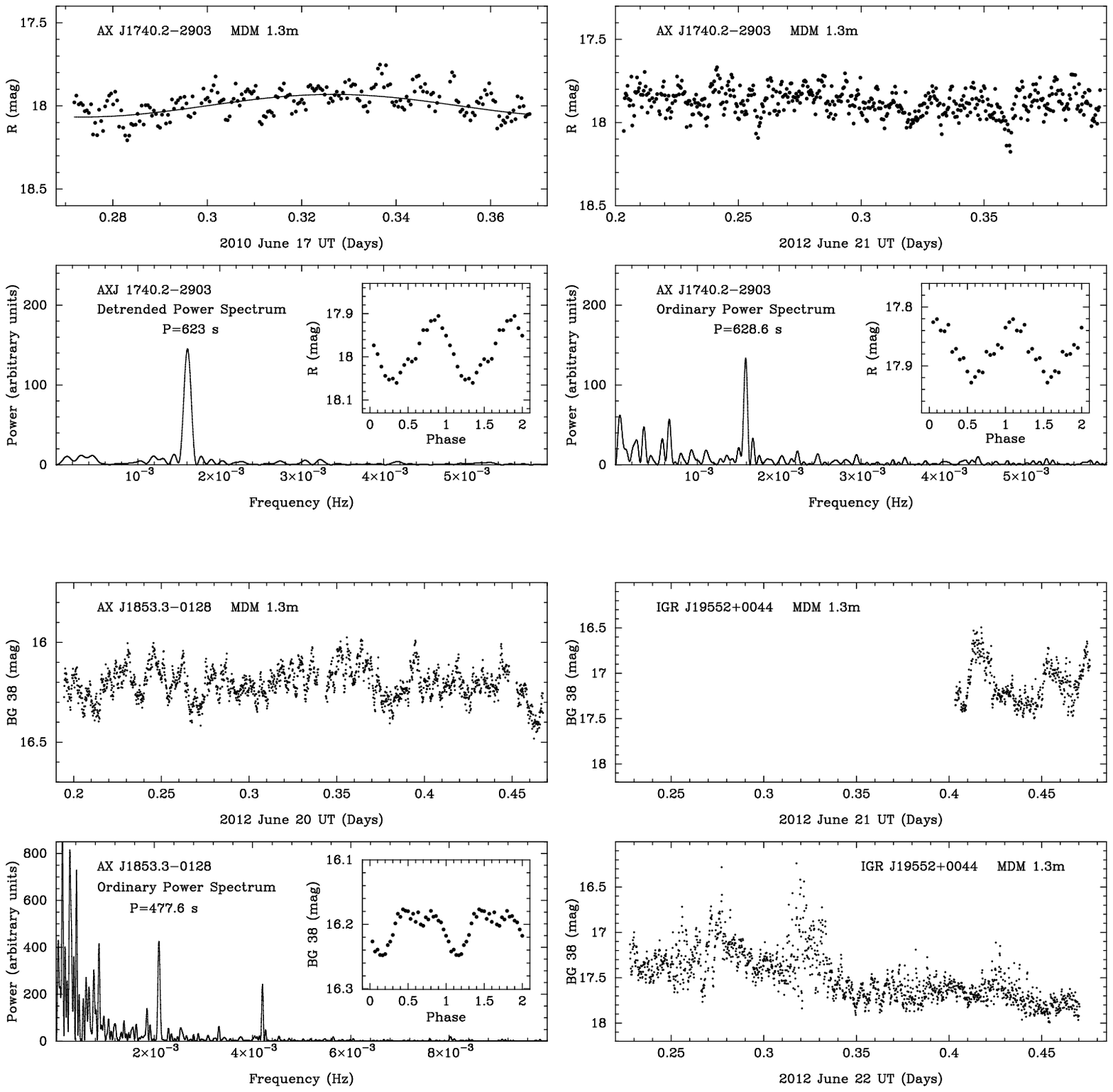}
}
\vspace{-2.7in}
\caption{Time-series photometry obtained on the 1.3m McGraw-Hill
telescope.  Individual exposures are 30~s in $R$
for \axa, and 10~s in BG38 for \axb\ and \igrd.  In cases where spin
periods are detected, the power spectrum, period,
and binned pulse profile (insets) are given in the panel below
the light curve.  In one case a broad trend (solid curve)
was subtracted from the data before calculating the ``detrended''
power spectrum.  Magnitude calibration in $R$ is from Landolt standard
stars.  In the BG38 filter, a rough calibration averaging $B$ and $R$
magnitudes from the USNO-B1.0 catalog is used.
}
\label{fig:photometry1}
\end{figure}

\begin{figure}
\vspace{-1.75in}
\centerline{
\includegraphics[scale=1.0,angle=0]{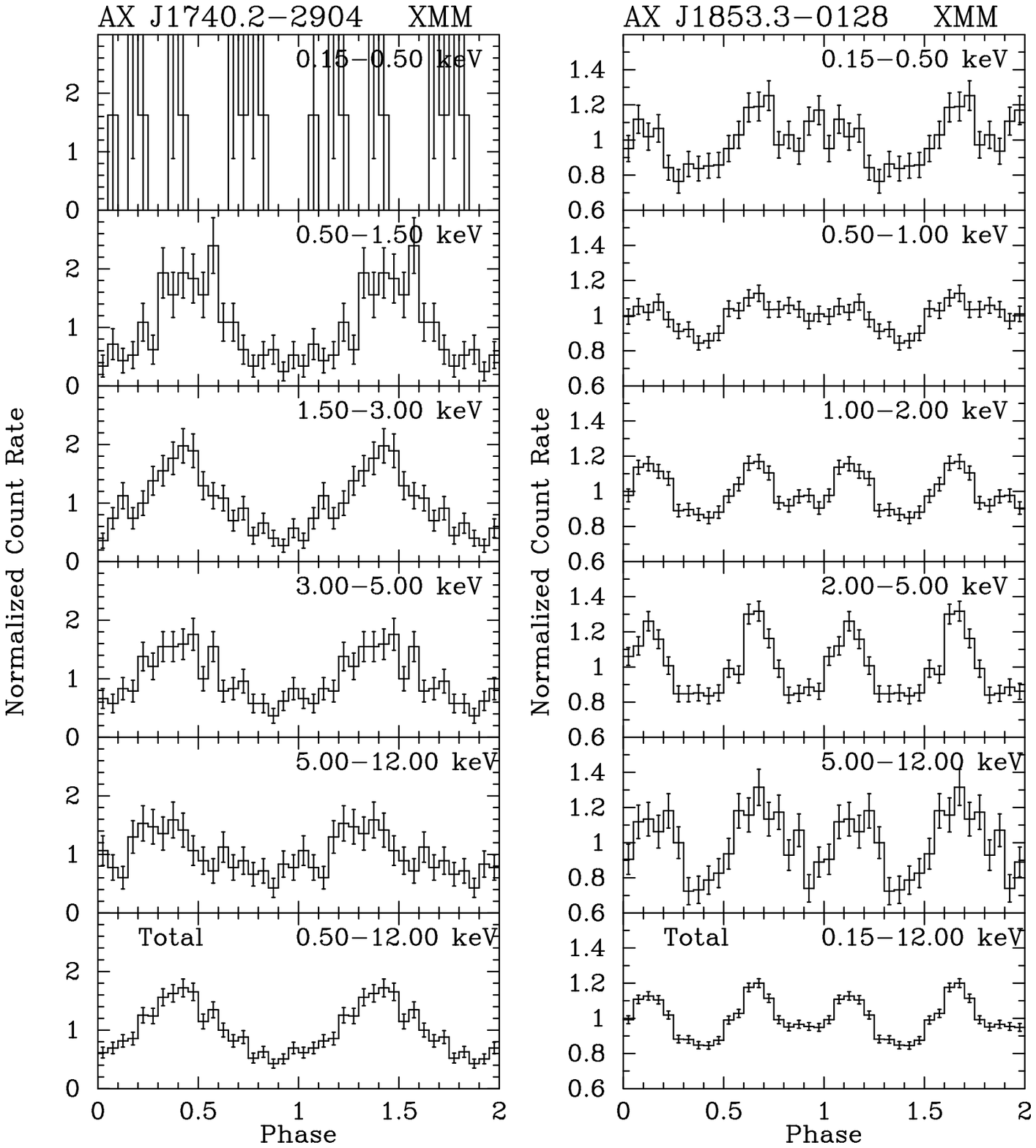}
}
\vspace{-2.25in}
\caption{Energy-dependent X-ray pulse profiles from \xmm\ observations.
Left: \axa\ folded on its period of 623~s, from an observation on
2005 September 29 (ObsID 03042201).
Right: \axb\ folded on its true 476~s period, from an observation
on 2004 October 25 (ObsID 02015003).
The pulse shape as a function of
energy changes from primarily a single, broad plateau at
energies $< 1$~keV (resembling the optical pulse), to
double peaked at high energies.
}
\label{fig:xraypulses}
\end{figure}

\citet{masetti09} identified the counterpart using a Swift
follow-up observation reported by \citet{landi08}, and 
obtained a spectrum. 
Our spectrum (Fig.~\ref{fig:montage2}) closely 
resembles theirs, with strong H$\alpha$ 
emission and no obvious indication of reddening.
\citet{masetti09} obtained direct images that gave $R=17.2\pm0.1$.  
Our direct images gave $V=16.9\pm0.2$, in 
rough agreement, and our spectra gave a 
synthesized $V=17.1$.

A spin period of $1690\pm18$~s was measured by \citet{ber12}
in the $B$ filter of the \xmm\ Optical Monitor, while the
X-ray power in the same observation is confined to the first
harmonic at $831.7\pm0.7$~s.  The latter is consistent with
$830.70\pm0.02$~s measured in \sw\ XRT data \citep{nic09}.
\citet{ber12} estimate
$P_{\rm orb}=8.5\pm0.2$ hr from
sidebands of the X-ray, and $P_{\rm orb}=6.6\pm0.3$ hr
by folding the X-ray light curves and looking for
a modulation.

The mean spectrum (Fig.~\ref{fig:montage2}) 
shows a blue continuum and strong 
emission lines.  
H$\alpha$ is double-peaked, with a weaker 
blue peak near $-270$~km~s$^{-1}$ and a stronger
red peak near $+330$~km~s$^{-1}$.  The line
profile is steep-sided, with a FWHM
$\approx1160$~km~s$^{-1}$.   A greyscale representation
of the H$\alpha$ line (Fig.~\ref{fig:swift0732trail}) shows
that the asymmetry of the red and blue 
persists throughout the orbital cycle, and is 
not caused by uneven sampling of an S-wave.
A very weak S-wave may be present, but it cannot
be traced across the center of the line.  

The radial velocities show a periodicity at
91.9 min, or 15.67 cycle~d$^{-1}$.  Because of the southerly declination 
and the time of year the data were taken, the 
velocities span only 5.7~hr of hour angle, so there
is some uncertainty in the daily cycle count.  
Running the Monte Carlo test against either one of the
flanking aliases gives a discriminatory power of
93\%; because there are two flanking aliases 
of nearly equal strength, the discriminatory power 
against the two is below 90\%.  The 
choice of radial-velocity period is therefore not entirely
reliable.  

The coherent 830/1660 s modulation shows clearly that
this is an IP.  A relatively
small number of such systems have orbital
periods short of the 2--3 hr ``gap''.
\citet{ritterkolb}, in their on-line catalog\footnote{available at
http://www.mpa-garching.mpg.de/RKcat/; we used 
Version 7.16}, list 50 objects as definite IPs;
of these, only four (V1025 Cen, HT Cam, DW Cnc, and EX Hya) have
orbital periods shorter than the 91.9 min
period found here; also, another 
object in the present sample, AX J1853.3$-$0128
(Section \ref{sec:ax1853}) has an 87-minute
period.  

While the radial-velocity period is very likely 
to be the orbital period, it is possible that it is
not.  In V455 And \citep{hasitall}, the emission-line
radial velocities follow persistent, but incoherent,
3.5-hour period, while $P_{\rm orb}$ is only
81.08 min.  The radial velocity of DW Cnc shows
two periods, 86.1 and 38.6 min, which are
probably the orbital period and the white-dwarf
spin period respectively \citep{pattdw, roddw}.
While the much longer $P_{\rm orb}$ candidates proposed
by \citet{ber12} are possible, neither of these
is definitive, and we think the radial velocity period
is more likely to be correct.  This object could use
further spectroscopy to untangle remaining cycle-count
uncertainties and determine whether our period is 
coherent over intervals longer than our 
week-long observing run -- if so, the case for the 
$P_{\rm orb} = 91$~min would be significantly stronger.

\subsection{\igrb}

The optical counterpart of this source was identified spectroscopically
by \citet{par08}, and \citet{masetti09} show the spectrum.
We have some direct images; one set of $BVI$ photometry images 
from 2010 June gives $V = 17.95$, $B-V =  +0.34$, and $V - I = 1.171$.  
\citet{masetti09} quote $R = 17.0$, in rough agreement.
\citet{ber12} found a spin period of $1820 \pm 2$ s in
\xmm\ data, and find a candidate $P_{\rm orb} = 4.2$~h from a weak,
presumed X-ray sideband.
The X-ray spin period appears to be reliable, but the evidence for an
orbital period is tentative.

The mean spectrum resembles the one published by
\citet{masetti09}; it shows H$\alpha$ with an emission
EW of $\sim 20$~\AA\ and a FWHM of $\sim 700$
km~s$^{-1}$.  The line profile is slightly
double-peaked, with the peaks at $\pm 150$ km~s$^{-1}$
from the line center.  The Na~I~D doublet is present
in absorption, with a combined EW $\sim 1.7$~\AA.
The Na~I~D line widths are unresolved, and the velocity
is constant, indicating that they are interstellar.
In addition, the DIB at $\lambda$5780 is present 
with EW $\sim 0.5$~\AA, 
which implies $E(B-V) \sim 0.8$ based on the  
correlations found by \citet{YuanLiuDIB},
or $A_V \sim 2.5$ assuming $A_V / E(B-V) = 3.1$.
\citet{masetti09} estimate a much lower
extinction, $A_V = 0.67$.  If $E(B-V)$ really
were as large as 0.8, our observed color
would imply $(B-V)_0 \sim -0.4$, which is too blue
for a novalike variable (see, e.g., the colors
compiled by \citealt{bruchengel94}), so the 
reddening is likely to be significantly less.   
As an illustration, adopting $E(B-V) = 0.4$,
or $A_V \sim 1.3$, gives $V_0 = 16.7$.  Novalike
variables have absolute magnitudes broadly similar
to dwarf novae in outburst -- $M_V \sim +4$ 
(\citealt{warner95}; see also \citealt{beuermann04}), 
which yields a distance around
3~kpc, much larger than the $\sim 320$ pc 
distance suggested by \citet{masetti09}, who
assumed $M_V \sim +9$.  While at least one IP
has an absolute magnitude this faint (EX Hya; 
\citealt{beuermann03}), we think \igrb\ is 
likely to be considerably more luminous.
Later, we find $P_{\rm spec} \sim 5.4$ h; at
that period the semi-empirical donor sequence
derived by \citet{kniggedonor} predicts that the 
donor should have spectral type M0 and $M_V = +8.2$.
We see no sign of a secondary contribution in our 
spectrum, which suggests $M_{V0} < 7$. 
Along this line of sight
($l = 19.\!^{\circ}45, b = -1.\!^{\circ}20$) the
reddening map of \citet{schlegel98} gives
a total Galactic extinction of $E(B-V) = 4.9$;
clearly, this object is in front of most of the 
dust.

The H$\alpha$ radial velocities give a period near 
5.4~hr.  If the frequency is 1~cycle~d$^{-1}$ higher,
the period is 4.39 hr, close to the period derived
by \citet{ber12}.  Our velocities span 
almost 6.5~hr of hour angle, and the \citet{tf85}
Monte Carlo test gives a discriminatory power of over
95\% for our alias choice.  Nonetheless, 
we cannot entirely rule out a cycle-count error. 
We obtained data on four observing runs with a 
total span of 778 days, so the period determination
suffers from multiple ambiguities in the longterm
cycle count as well as a daily ambiguity.  
The period uncertainty in Table~\ref{tab:parameters}
is based on the lengths of the individual observing
runs.

\subsection{\axb}
\label{sec:ax1853}

The periodic nature of this source in a 20~ks \xmm\ observation 
taken on 2004 October~25 was noted
by \citet{mun08} and \citet{hui12}.  Both authors concluded that the
period is 238~s, supporting an IP classification.
However, detailed inspection of the pulsed X-ray light curve
reveals that the true period is twice this value.  Figure~\ref{fig:xraypulses}
shows the energy-dependent pulse profiles folded at our derived period
of $476.0 \pm 0.2$~s.  Even though most of the power falls at the first 
harmonic because the hard X-ray pulse is double peaked, it is apparent
that the two minima of the light curve are of unequal heights, the
difference growing more prominent toward low energies until
the soft X-ray pulse ($E<1$~keV) has a single broad plateau
and barely two maxima.  Folding at half the correct period
was largely responsible for the small modulation at the
lower energies in the pulse profiles displayed in \citet{hui12}.

We are not aware of any prior optical work on \axb.
We identified its optical counterpart from the \xmm\
position, and obtained time-series photometry on the 1.3m McGraw-Hill
on 2007 June 19 through a BG38 filter.  This 3-hr run, while displaying
flickering typical of a CV, did not reveal a periodic signal.
However, a repeat of this observation in a 6.5-hr run on 
2012 June 20 (Fig.~\ref{fig:photometry1}) clearly
confirms the 476~s period and the IP classification.
We find $P = 477.6 \pm 1.0$~s. Unlike the X-rays, the
majority of the optical power (in the broad BG38 filter)
is in the fundamental, although the first harmonic is also prominent.
The binned light curve has an amplitude of only 0.07~mag
peak-to-trough, but its pulse shape is strikingly similar 
to that of the soft X-rays, a broad, flat plateau with a slight
dip in the middle.
It is possible that the double-peaked X-ray light curve is indicating
emission from both poles because the peaks are $180^{\circ}$ apart.
However, in the optical light curve the peaks are not $180^{\circ}$ apart.

The emission lines in the mean spectrum 
(Fig. \ref{fig:montage2}) are unusually strong
and quite broad;
H$\alpha$ has an EW of 230~\AA\ and a FWHM
$\sim 1350$ km s$^{-1}$.  All the emission
lines are slightly double-peaked; in 
H$\alpha$, the blue peak is at $-360$ km s$^{-1}$
and the red at $+200$ km s$^{-1}$.  The 
red peak is slightly stronger than the blue peak.
The H$\alpha$ radial velocities are modulated
on an 87-minute period, with no ambiguity in
daily cycle count.  In the optical, 
\axb\ appears to be a near-twin of 
\igra\ (Section \ref{sec:igr1817}).

\subsection{\igrc}

\citet{steeghs08} identified this as a cataclysmic
binary using the IPHAS H$\alpha$ survey \citep{withamiphas} and a Chandra
localization \citep{tomsick08,tomsick09}.
\citet{masetti09} show a spectrum.   \citet{evans08}
detected pulsations at 938.6(+5.6,$-$5.9) s in
Swift-XRT data, showing that this is a
DQ Her-type system.  They also detected some
evidence for a period at $\sim$~16500(+1900, $-$1500) s,
which they suggested might be an orbital period. 

Our mean optical spectrum shows a red-sloping 
continuum with strong, single-peaked emission  
lines -- H$\alpha$ has EW~$\sim$~80~\AA, and
FWHM $\sim$ 900~km~s$^{-1}$.  He~II~$\lambda$4686
is about half the strength of H$\beta$, and 
He~II~$\lambda$5411 is present as well.  

Most of our optical spectra were obtained on 
adjacent nights in 2008 September, and the
remainder on three nights in 2009 June.  The
data define an unambiguous 3.4-h orbital period
(Table~\ref{tab:parameters}), but 
with an unknown cycle count in the 280-day
gap between observations.  Our period amounts
to 12416 $\pm$ 35 s, so it appears that the
16500-s period detected by \citet{evans08}, if 
persistent, is unrelated to the orbit.

\begin{figure}
\centerline{
\includegraphics[scale=0.82,angle=0]{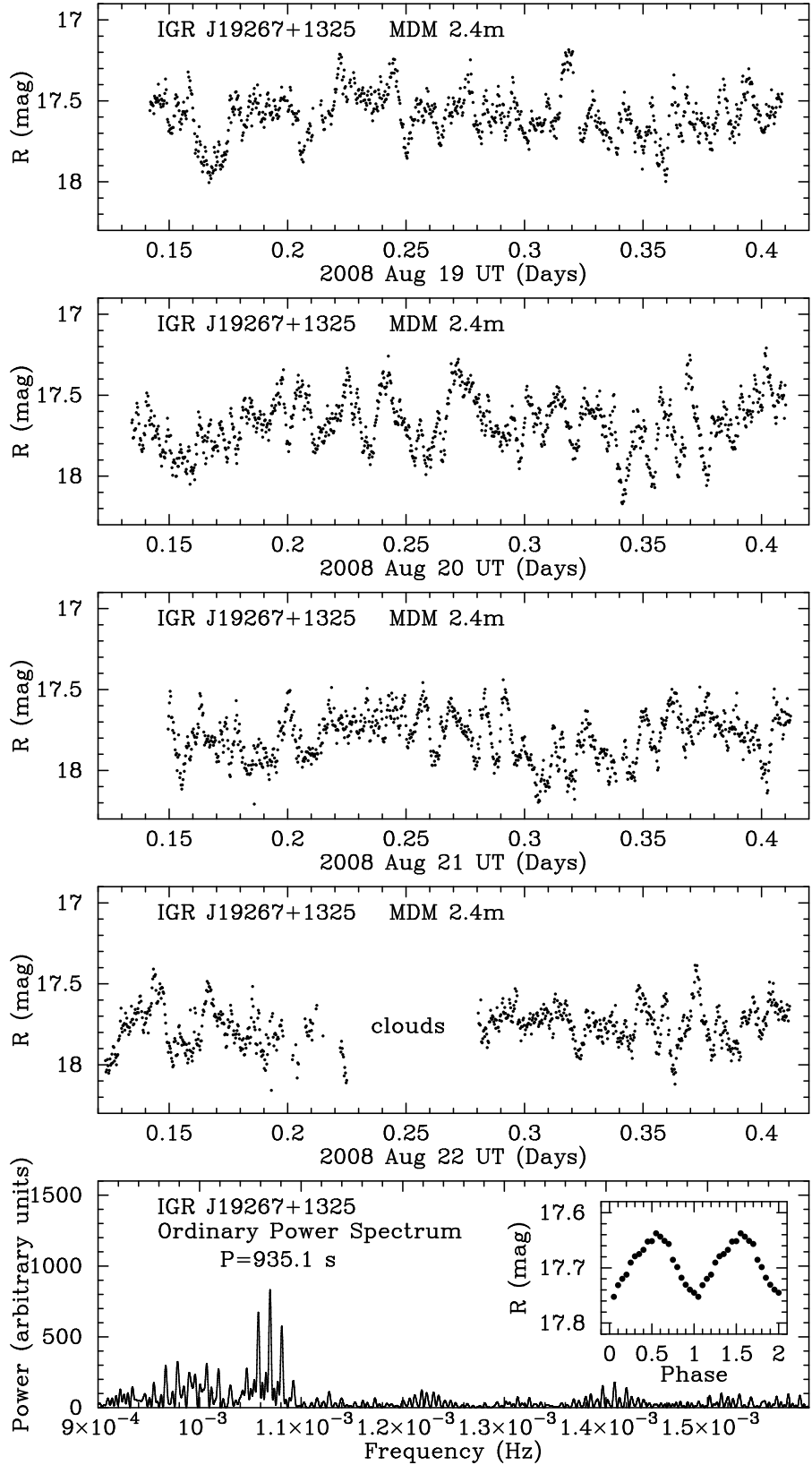}
}
\caption{Time-series CCD photometry of \igrc\
on four consecutive nights.  Individual exposures are 10~s
in $R$.  The coherent power spectrum of the four nights and the
folded light curve are given in the bottom panel.  The measured
spin period is $935.1\pm 0.2$~s.
}
\label{fig:photometry2}
\end{figure}

We obtained time series photometry of \igrc\ in the $R$ filter
on four consecutive nights in 2008 August (Fig.~\ref{fig:photometry2}).
A spin period is detected on each night individually, 
and a coherent power spectrum of all four nights gives $P=935.1\pm0.2$~s,
in agreement with the X-ray period of \citet{evans08}.  On one
night, August 21, the strongest peak was at 1009~s, which
could possibly be the sideband, or beat between the spin
and the orbit.

\subsection{\igrd}
\label{sec:igr1955}

\begin{figure}
\plotone{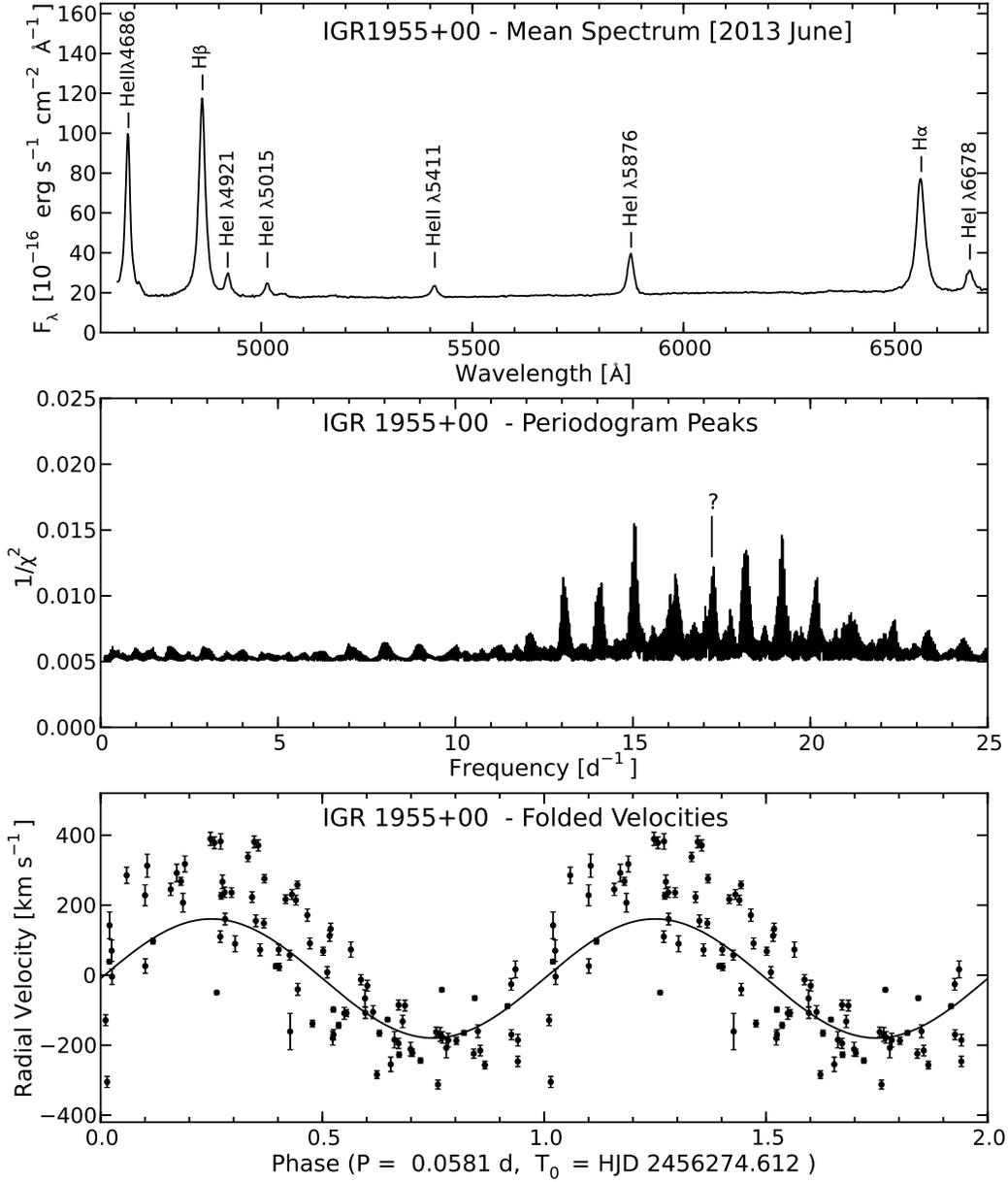}
\caption{
Top panel: Mean spectrum of \igrd\ from 2013 June. 
Middle panel: Period search of H$\alpha$ radial velocities of 
\igrd.  The full periodogram is not shown; rather, the curve
connects local maxima.   Lower panel: H$\alpha$ radial velocities 
folded on the adopted period, which is uncertain. 
A sinusoid fitted at this period (Table~\ref{tab:parameters}) is also
shown.
}
\label{fig:igr1955vels}
\end{figure}

\citet{mas10} give a finding chart and a spectrum 
showing unusually strong Balmer lines together
with He~I and some He~II emission.  Our spectra appear 
generally similar 
to theirs; the top panel of Fig.~\ref{fig:igr1955vels}
shows the mean of our 2013 June spectra.  HeII $\lambda$4686
and $\lambda$5411 appear strongly.

We have radial velocities from three observing
runs late in 2012, one in 2013 February, and limited data from
2013 June.  Unfortunately, we could not determine an 
unambiguous rough orbital period.
The most extensive and `cleanest' data are from two nights
in 2012 October, which
favored a radial-velocity frequency near 17.17 cycle d$^{-1}$
(or 83.9 min),
with aliases spaced by 0.5 cycle d$^{-1}$, reflecting a 
two-day gap between the two nights of observation.  
The 1-day baselines in the remainder of the data eliminate every other 
one of these aliases,
but when all the data are analyzed together there is strong
scatter in the velocity curve, evidently reflecting changes
in the state of the source from run to run.  Fig.~\ref{fig:igr1955vels}
(middle panel)
shows the periodogram of all our velocities; it 
is badly affected by a combination
of large variations in the source and limited sampling in the
various observing runs.  We very tentatively adopt a 
frequency (marked by a question mark) that is consistent
with the 2012 October data and which yields a folded
velocity fit (lower panel of Fig.~\ref{fig:igr1955vels}) 
that is roughly consistent with the remainder of the velocities.  
Although this is very crude, the velocities clearly
indicate a period shortward of the 2-3 hour period
gap.  

\begin{figure}
\plotone{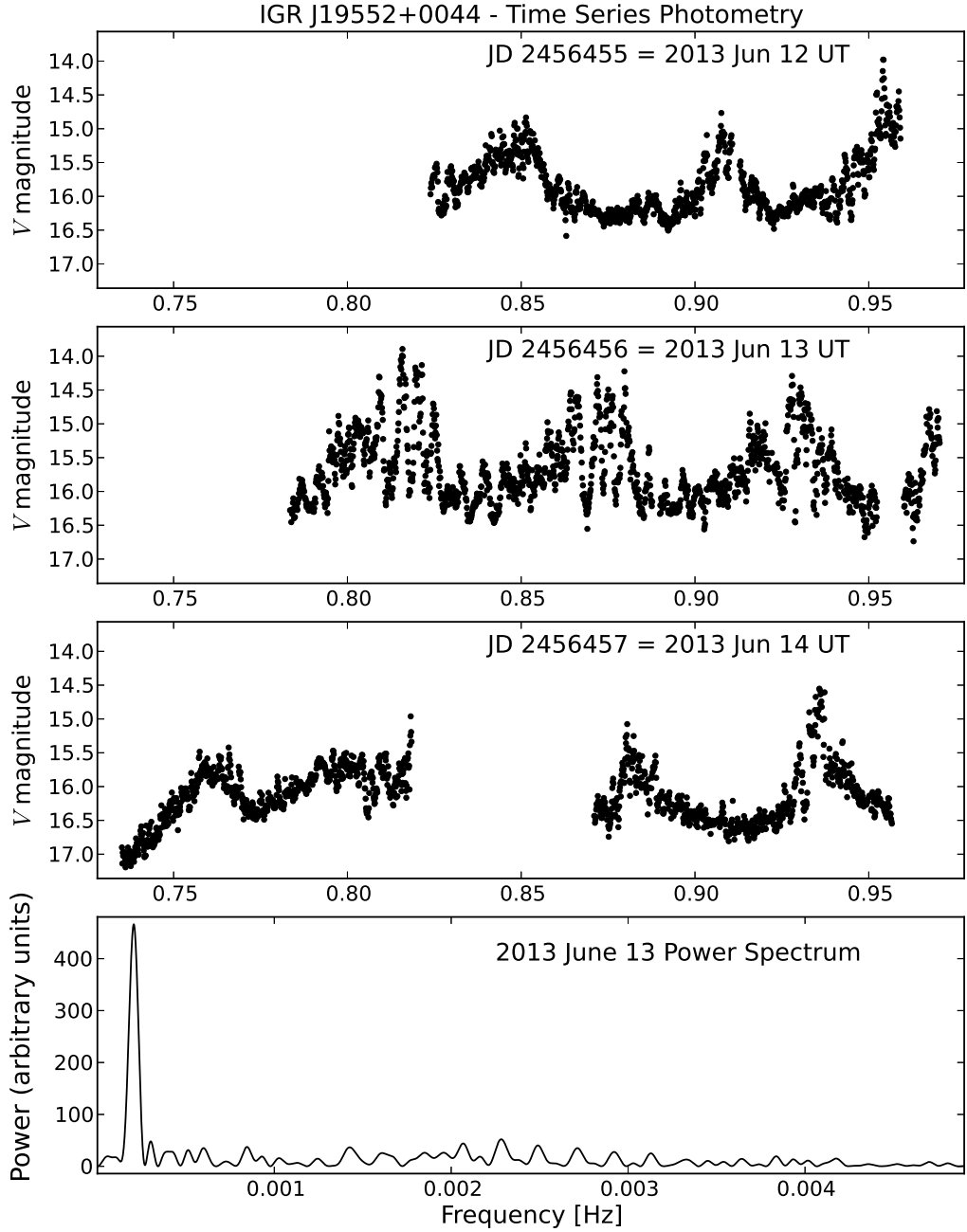}
\caption{
Time-series photometry of \igrd\ obtained in 2013 June.  
For the top three panels, the horizontal axes show fractional
days on the dates indicated.  The lower panel shows a power
spectrum for June 13; the peak at low frequency corresponds to the
$\sim 81$ min modulation evident in the light curve.  The 
absence of other peaks indicates that the
large-amplitude flickering on that night 
does not arise from a coherent periodicity.
}
\label{fig:igr1955phot}
\end{figure}

We obtained brief photometric time series 
of \igrd\ 
in 2012 June (Fig.~11) which showed no
evidence of a coherent pulsation,
although the star flickered by over 1 magnitude
on a time scale of minutes.
On three nights in 2013 June we obtained 
more extensive differential photometry 
using the Andor camera and a $V$ filter.  To convert
the 2013 June time series
to approximate standard $V$ magnitudes, we used
the SDSS Data Release 9 to estimate 
$V = 16.94$ for the main comparison star, which
lies 8 arcsec west and 39 arcsec north of \igrd . 
The light curves (Fig.~\ref{fig:igr1955phot})
show an
approximately periodic brightening on a timescale
similar to our adopted radial-velocity period.  A sinusoidal
fit to these data gives a best period
near 0.0565 d ($\sim 81$ min), or 17.7 cycle d$^{-1}$.  Our 
best radial velocity period amounts to 17.23
cycle d$^{-1}$, so the photometric modulation
does not appear to be coherent on the 
radial-velocity period; constraining the
radial-velocity period to a range consistent with the
apparent photometric period results in very poor fits.

The photometry from 2013 Jun 13 UT also shows
flickering with a timescale of 5 to 10 minutes
and an amplitude of up to $\sim 2$ magnitudes.
This is not an artifact, as it is obvious from
inspection of the original images.  
A power-spectrum analysis of that night's data
(Fig.~\ref{fig:igr1955phot}, lower panel)
showed no coherent signal consistent with the 
flickering time scale.

It is not obvious how to classify this object.
The absence of any coherent pulse is consistent with 
an AM Her star, or polar.  However, in our limited data
we do not see strong, asymmetric line wings or
sharp moving features in our spectra (as we see
in \swc; see Section \ref{sec:swift2218}).  Time-resolved
polarimetry might clarify the classification.

\subsection{\swc}
\label{sec:swift2218}

This was detected as a hard X-ray source in the \sw\ BAT
\citep{cus10b,bau12}, and localized by the \sw\ XRT in
observations on 2009 August 2 and 4.  It is coincident
with 1RXS J221832.8+192527.  Its highly variable \sw\
X-ray light curves are shown in Figure~\ref{fig:swiftxrt}.
We present a finding chart
from the \sw\ UVOT in Figure~\ref{fig:swiftuvot}, which
shows a UV-bright counterpart.
We reported briefly on optical observations
of \swc\ in \citet{tho09}, and give more details here.  
We are not aware of any other optical work on this source.

\begin{figure}
\epsscale{0.95}
\plotone{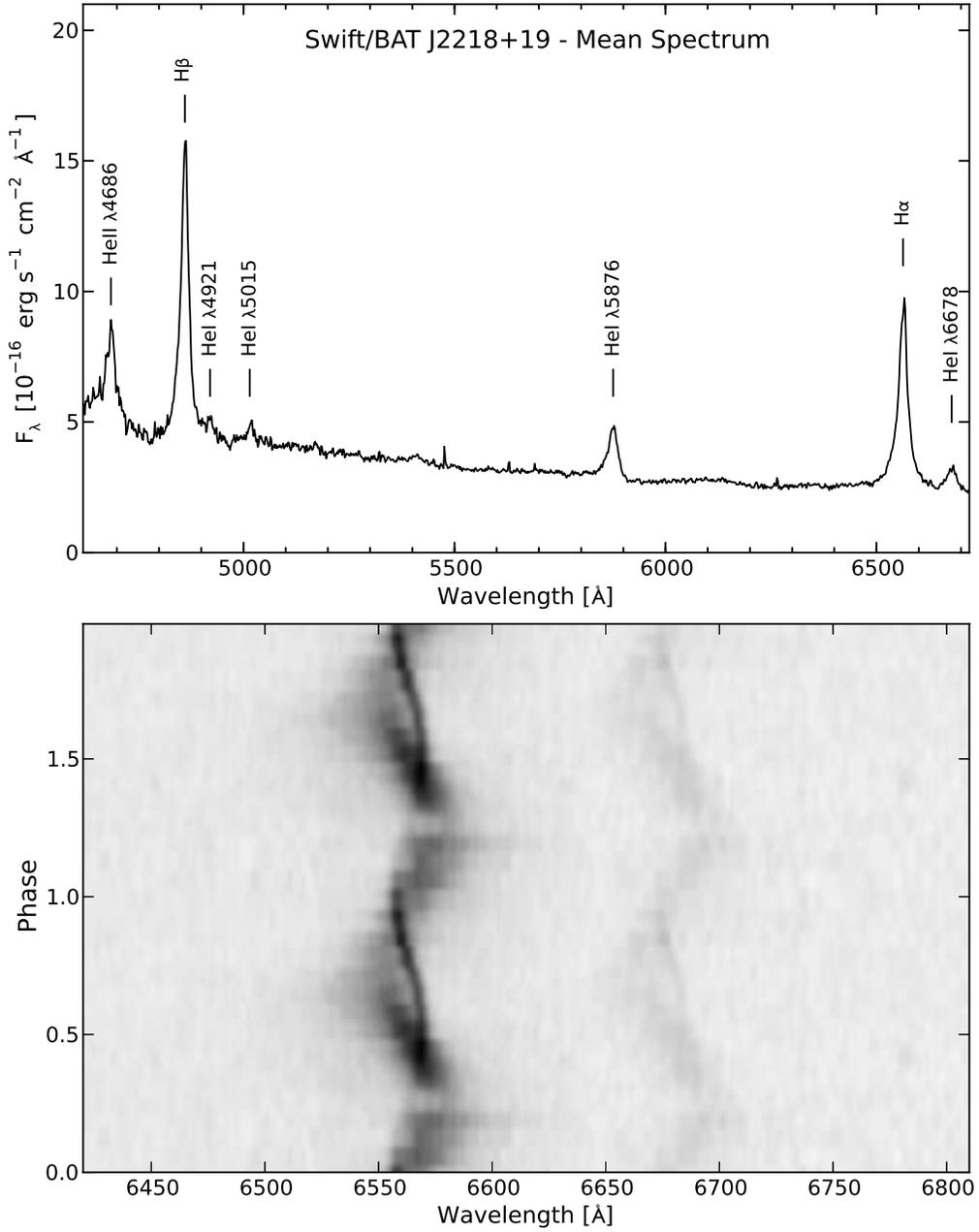}
\caption{
Upper panel: Mean spectrum of \swc.  The more prominent
emission lines are identified.  Lower panel: A greyscale representation 
of the region of the H$\alpha$ and He~I~$\lambda$6678 emission lines,
plotted as a function of orbital phase.  Phase increases from the
bottom, and two cycles are shown for continuity.  The spectrum
is represented as a negative (i.e., larger fluxes are shown as 
darker pixels).  The phase is computed from the values of 
$T_0$ and $P_{\rm spec}$ listed in Table~\ref{tab:parameters}.  
See Section \ref{sec:swift0732} for procedural details.
}
\label{fig:swift2218fig}
\end{figure}

The average spectrum (Fig.~\ref{fig:swift2218fig}) shows strong, 
broad emission lines, and He~II $\lambda4686$ is clearly
detected.  The amplitude of the H$\alpha$ radial velocity modulation 
is large, and the 129-min period is determined without ambiguity.
The large velocity amplitude, and extensive line wings, suggest that
this is a polar, or AM Herculis star, in which much of the 
emission comes largely from a synchronously-rotating, 
magnetically-channelled accretion column.  

The lower panel of Fig.~\ref{fig:swift2218fig} shows a greyscale 
phase-averaged spectrum synthesized from our spectra, prepared
using the procedure discussed in Section \ref{sec:swift0732}.
The image reveals two distinct emission-line components:
one of them is diffuse and has a large velocity amplitude, 
and the other is sharper, has a lower amplitude, and is 
visible over only part of the phase.  The diffuse component 
very likely arises in an accretion column, and
the sharp component is probably radiation reprocessed
on the side of the secondary star that faces the white dwarf.
The sharp component comes into view around its maximum 
positive velocity, and fades from view near its minimum velocity;
this is just as expected from a heated face.
The behavior of the emission-line profiles with phase 
closely resembles that of known polars \citep{schwope99}, so 
we conclude that \swc\ is another example of this type.

Using the single-trail image, we estimated radial velocities of the
sharp component by eye and fit them with a sinusoid, which had
$T_0 = \hbox{HJD } 2455070.7050(7)$ and $K_2 = 282(15)$ km s$^{-1}$.
The epoch $T_0$ falls 0.26 cycle later than the $T_0$ listed in 
Table \ref{tab:parameters}; it corresponds to the blue-to-red
transition of the sharp component, so it very likely represents the 
inferior conjunction of the secondary star.  The epoch in 
Table \ref{tab:parameters} is based on ``whole-line'' measurements,
which do not have a simple physical interpretation but which do
define the period well.  

If the masses of the component stars are
broadly typical for CVs of this period, and the
sharp component traces the motion of the secondary, then the orbital
inclination $i$ is unremarkable.  For illustration, taking
$M_1 = 0.75$ M$_\odot$ and $M_2 = 0.2$ M$_{\odot}$ (using
\citealt{kniggedonor} as a guide), our $K_2$ implies
$i = 50^{\circ}$, so eclipses are unlikely.

\section{Conclusions}
\label{sec:conclusions}

The selection of CVs studied here does not comprise
a well-defined ``sample'' in any quantitative sense.
However, selected on the basis of hard X-ray detection
and variability, they lead to the expected result that
at least eight of the ten are magnetic CVs, of which seven
are of the IP class, which is the
dominant CV population in hard X-ray surveys.
Their spin and orbit periods are within the ranges
previously observed for members of their classes.
Here we summarize the most interesting conclusions
about the individual objects, and offer recommendations
for further work on some of these stars.

We obtained firm orbital periods for eight of the ten objects, 
ranging from 87~min to 9.38~hr. 
Two of the IPs studied here are somewhat unusual in having short
orbital periods, \igra\ (91.9~min) and \axb\ (87~min).
\igra\ has the unusually large value of $P_{\rm spin}/P_{\rm orb}=0.3$,
and is an exception to the observation of \citet{sca10}
that all hard X-ray-detected IPs have $P_{\rm spin}/P_{\rm orb}<0.1$.
The spin period of \axb, 476~s, is twice the value previously
published on the basis of X-ray data.  A careful analysis reveals
its fundamental period in X-rays as well as in optical.

We obtained the first evidence for the magnetic nature of
\rxs\ by detecting a persistent 1218~s period, and
confirmed the IP nature of \igrc\ by detecting its 935~s
period in the optical over four nights. 

The previously debated nature of \axa\ is resolved in favor of
an IP by showing that its 626~s X-ray (spin) period is also present
in the optical, and by discovering its 5.7~hr orbital period.
Together, these are typical results for a CV, but the 626~s period
is too long for a neutron star in a low-mass X-ray binary.  The
assumed WD nature of the compact object could be further tested
with campaign of coherent timing of its rotation; long-term constancy
would be expected for a WD, but not for a neutron star.

Our study identified two new CVs, \swb\ and \swc.  The latter
is clearly an AM~star with an orbital period of 129~min,
while \swb\ is of an uncertain type in which the
secondary star is prominent in the spectrum and displays
the 9.384~hr orbit. We estimated the secondary star as
K4$\pm2$ at a distance of 900(+190,--150)~pc.
At $V\sim14.5$, \swb\ is an easy target for additional
searches for a spin period.

\igrd\ has a short but ambiguous orbital period, probably
around 84~min.  Its high-amplitude flickering with no 
coherent period,
and large radial velocity amplitude, suggest that it may
be an AM~Her star.  Follow-up studies, especially polarimetry,
could better determine its nature.

\acknowledgments

We gratefully acknowledge support from NSF grants 
AST--0708810 and AST--1008217.  We also thank Hwei-Ru Ong
and Arlin Crotts for obtaining some of the optical spectra of \igrd.
The \xmm\ observations of \axa\ and \axb\ were conducted
by Eric Gotthelf, who first found their periodic signals.
\xmm\ is an ESA science mission with instruments
and contributions directly funded by ESA Member States and NASA.

\clearpage


\clearpage

\begin{deluxetable}{lccrclccc}
\tablecolumns{9}
\tablewidth{0pt}
\tablecaption{Stars Observed}
\tablehead{
\colhead{Name} &
\colhead{R.A.} &
\colhead{Decl.} &
\colhead{$V$} & 
\colhead{Ref\tablenotemark{a}} & 
\colhead{Data\tablenotemark{b}} & 
\colhead{Class\tablenotemark{c}} &
\colhead{$P$} &
\colhead{Ref\tablenotemark{d}}\\
\colhead{} &
\colhead{(h m s)} &
\colhead{($^\circ$ $'$ $''$)} & 
\colhead{} & 
\colhead{} & 
\colhead{} & 
\colhead{} &
\colhead{(s)} &
\colhead{}
}
\startdata
\rxs & 04 57 08.32 & +45 27 50.0 &  17.5 & S & S,T & DQ & 1218.7(5) & 1 \\
\swa\tablenotemark{e} & 07 32 37.71 & $-$13 31 08.3 & 15.7 & D & S & DQ & 512.42(3) & 2 \\
\swb & 07 46 17.11 & $-$16 11 27.7 & 14.5 & A & S,X & N  \\
\axa & 17 40 16.10 & $-$29 03 38.1 &  19.1 & S & S,X,T & DQ & 628.6(2.3) & 1 \\
\igra & 18 17 22.18 & $-$25 08 42.6 &  16.9 & D & S & DQ & 1663.4(1.4) & 3 \\
\igrb & 18 30 49.94 & $-$12 32 19.1 &  17.9 & D & S & DQ & 1820(2) & 3 \\
\axb & 18 53 30.60 & $-$01 28 15.9 &  16: & B & S,X,T & DQ & 477.6(1.0) & 1\\
\igrc & 19 26 27.00 & +13 22 04.9 &  18: & I & S,T & DQ & 935.1(2) & 1 \\
\igrd & 19 55 12.47 & +00 45 36.6 &  16:\tablenotemark{f} & B & S,T & AM? \\
\swc & 22 18 32.76 & +19 25 20.5 &  17.5  & B & S,X & AM
\enddata
\tablecomments{Coordinates are for J2000.0 (ICRS), either
from the PPMXL catalogue \citep{ppmxl} or derived
from astrometric fits to our own images referred to the
UCAC-3 \citep{ucac3}.  Estimated uncertainty is   
$\pm 0.\!^{\prime\prime}2$.}
\tablenotetext{a}{Source of approximate $V$ magnitude: S = our
spectrophotometry; D = our direct image; A = APASS \citep{apass}, 
as listed in the UCAC-4 \citep{ucac4}; 
B = interpolated from Schmidt-plate magnitudes in USNO-B1.0;
I = extrapolated from IPHAS, $r' = 17.7$ \citep{steeghs08}.
\citep{mon03}, as listed in PPMXL \citep{ppmxl}.}
\tablenotetext{b}{Types of data presented here; 
S = time-resolved spectroscopy; T = time series
photometry; X = X-ray light curve.}
\tablenotetext{c}{Classifications are: N = novalike
variable (pulsations not confirmed); DQ = DQ Her
star or IP (evidence for pulsations); 
AM = AM Her star or polar.}
\tablenotetext{d}{Reference for period $P$, presumed to be the spin period:
(1) this paper, optical; (2) \citealt{patterson06}, optical; 
(3) \citealt{ber12}, X-ray.}
\tablenotetext{e}{Now named V667 Puppis}
\tablenotetext{f}{Catalogued magnitudes of this object show 
variability by $\sim 2$ mag.}
\label{tab:objects}
\end{deluxetable}

\begin{deluxetable}{lrccrr}
\tablecolumns{6}
\tablewidth{0pt}
\tablecaption{Radial Velocities}
\tablehead{
\colhead{Star\tablenotemark{a}} &
\colhead{Time\tablenotemark{b}} &
\colhead{$v_{\rm abs}$} &
\colhead{$\sigma$} & 
\colhead{$v_{\rm emn}$} &
\colhead{$\sigma$} \\
\colhead{} &
\colhead{} &
\colhead{(km s$^{-1}$)} &
\colhead{(km s$^{-1}$)} &
\colhead{(km s$^{-1}$)} &
\colhead{(km s$^{-1}$)}
}
\startdata
RX J0457 & 55128.9305  &  \nodata & \nodata &  $   -3$ & $   7$ \\
RX J0457 & 55128.9373  &  \nodata & \nodata &  $    4$ & $   5$ \\
RX J0457 & 55128.9578  &  \nodata & \nodata &  $   -8$ & $   5$ \\
RX J0457 & 55128.9654  &  \nodata & \nodata &  $  -19$ & $   6$ \\
RX J0457 & 55129.7049  &  \nodata & \nodata &  $   39$ & $   8$ \\
RX J0457 & 55129.7153  &  \nodata & \nodata &  $   35$ & $   6$ \\
RX J0457 & 55129.7254  &  \nodata & \nodata &  $   20$ & $   5$
\enddata
\tablenotetext{a}{Abbreviated star name.}
\tablenotetext{b}{Heliocentric Julian Date of mid-exposure minus 2400000; the 
time base is UTC.}
\tablecomments{Table \ref{tab:velocities} is published in its entirety in the electronic 
edition of The Astronomical Journal, A portion is shown here for guidance regarding its form and content.}
\label{tab:velocities}
\end{deluxetable}

\clearpage

\begin{deluxetable}{lllrrcc}
\tablecolumns{7}
\footnotesize
\tablewidth{0pt}
\tablecaption{Fits to Radial Velocities}
\tablehead{
\colhead{Data set} & 
\colhead{$T_0$\tablenotemark{a}} & 
\colhead{$P_{\rm spec}$} &
\colhead{$K$} & 
\colhead{$\gamma$} & 
\colhead{$N$} &
\colhead{$\sigma$\tablenotemark{b}}  \\ 
\colhead{} & 
\colhead{} &
\colhead{(d)} & 
\colhead{(km s$^{-1}$)} &
\colhead{(km s$^{-1}$)} & 
\colhead{} &
\colhead{(km s$^{-1}$)}
}
\startdata
\rxs & 55221.924(6) & 0.257819\tablenotemark{c} &  44(6) & $-25(5)$ & 60 &  21 \\ 
\hspace{0.1in}(alternate) & 55221.875(5) & 0.200424\tablenotemark{c} &  46(6) & $-27(5)$ & 60 &  21 \\[1.2ex] 
\swa\ &  53812.7591(14) & 0.2338(2) &  149(5) & $ 20(4)$ & 50 &  18 \\[1.2ex]
\swb\ (absn) & 55279.663(3) & 0.391012(11) &  90(4) & $ 53(3)$ & 47 &  13 \\
\hspace{0.1in}(emission) & 55279.524(9) & 0.39097(3) &  54(6) & $ 35(5)$ & 48 &  21 \\
\hspace{0.1in}(combined) & \nodata & 0.391006(10) & \nodata & \nodata & \nodata  \\[1.2ex]
\axa\  & 55736.8052(19) & 0.2384(5) &  110(6) & $ 42(4)$ & 29 &  18 \\[1.2ex]
\igra\  & 55365.6931(17) & 0.06382(8) &  47(8) & $ 44(6)$ & 35 &  22 \\[1.2ex]
\igrb\  & 55734.698(2) & 0.2239(5) &  104(9) & $-6(5)$ & 49 &  26 \\[1.2ex]
\axb\  & 56078.7312(9) & 0.06058(7) &  42(4) & $-53(3)$ & 30 & 12 \\[1.2ex]
\igrc\ (2008 Sep) & 54712.6507(17) & 0.1455(8) &  80(6) & $ 16(4)$ & 45 &  16 \\ 
\hspace{0.1in}(2009 Jun)  & 54996.885(3) & 0.1432(4) &  72(6) & $ 25(5)$ & 13 &  14 \\ 
\hspace{0.1in}(combined) & 54712.6504(20) & 0.1437(4)  &  79(7) & $ 17(5)$ & 45 &  19 \\[1.2ex]
\igrd & 56274.6121(20) & 0.058051 :\tablenotemark{d} & 170(33) & $-9(24)$ & 97 & 112 \\[1.2 ex]
\swc  & 55070.6819(9) & 0.08996(9) & 344(26) & $ 25(16)$ & 32 &  71 \\[1.2ex]
\enddata
\tablecomments{Parameters of least-squares fits to the radial
velocities, of the form $v(t) = \gamma + K \sin\,[2 \pi(t - T_0)/P_{\rm spec}]$.}
\tablenotetext{a}{Heliocentric Julian Date minus 2400000.  The epoch is chosen
to be near the center of the time interval covered by the data, and
within one cycle of an actual observation.}
\tablenotetext{b}{Root-mean-square residual of the fit.}
\tablenotetext{c}{The two periods reflect different choices of daily cycle count, and 
the excessively precise periods tabulated here reflect in turn an arbitrary choice of 
cycle count
between observing runs.  For each choice of daily cycle count, we estimate the
uncertainty in the gross period to be 0.008~d.}
\tablenotetext{d}{This period was used in the fit, but it is highly
uncertain
-- see Section \ref{sec:igr1955} and Figure~\ref{fig:igr1955vels}.}
\label{tab:parameters}
\end{deluxetable}

\end{document}